\documentclass[10pt,a4paper,twocolumn]{article}

\usepackage{epsf}
\usepackage{amssymb}
\usepackage{latexsym}
\usepackage{graphicx}
\usepackage{parskip}
\usepackage[english]{babel}
\usepackage{amsmath}

\usepackage[affil-it]{authblk}
\usepackage{sectsty}

\sectionfont{\large}
\subsectionfont{\normalsize}

\newcommand{\EX}{\mathtt{E}}
\newcommand{\VA}{\mathtt{Var}}
\newcommand{\elemPt}{\mathtt{b}}

\newcommand{\PtsS}{\mathcal{P}}
\newcommand{\Pt}{\mathtt{A}}

\begin{document}

\title{Percentiles of sums of heavy-tailed random 
variables:\\Beyond the single-loss approximation.}

\author[1]{Lorenzo Hern\'andez \thanks{lorenzo.hernandez@qrr.es}}
\author[1]{Jorge Tejero \thanks{jorge.tejero@qrr.es}}
\author[2]{Alberto Su\'arez \thanks{alberto.suarez@uam.es}}
\author[1,3]{Santiago Carrillo-Men\'endez \thanks{santiago.carrillo@uam.es}}
\affil[1]{Quantitative Risk Research S.L. Madrid, Spain}
\affil[2]{Computer Science Dpt., Universidad Aut\'onoma de  Madrid. Madrid, Spain}
\affil[3]{Mathematics Department, Universidad Aut\'onoma de Madrid. Madrid, Spain}

\maketitle

\begin{abstract}
A perturbative approach is used to derive approximations of arbitrary order to estimate high percentiles of sums of positive independent random variables that exhibit heavy tails. Closed-form expressions for the successive approximations are obtained both when the number of terms in the sum is deterministic and when it is random. The zeroth order approximation is the percentile of the maximum term in the sum. Higher orders in the perturbative series involve the right-truncated moments of the individual random variables that appear in the sum. These censored moments are always finite. As a result, and in contrast to previous approximations proposed in the literature, the perturbative series has the same form regardless of whether these random variables have a finite mean or not. 
For high percentiles, and specially for heavier tails, the quality of the estimate improves as more terms are included in the 
series, up to a certain order. 
Beyond that order the convergence of the series deteriorates. Nevertheless, the approximations obtained 
by truncating the perturbative series at 
intermediate orders are remarkably accurate for a variety of 
distributions in a wide range of parameters. 


\end{abstract}

\section{Introduction}
\label{intro}
In this article we derive accurate closed-form approximations 
for high percentiles of sums of positive 
independent identically distributed 
random variables (iidrv's) with heavy tails. 
This is an important computational task in applications such 
as wireless communications \cite{nadarajah_2008_review},  
workload process \cite{cohen_1972_tail,fay++_2006_modeling}
and in the quantification of risk in insurance 
and finance \cite{embrechts++_1997_modelling,mcneil2005}. 
A particularly important application in finance
is the quantification of operational risk
\cite{frachot++_2001_loss,embrechts++_2003_quantifying,panjer_2006_operational,carrilloMenendez+suarez_2012_robust}.

There are several numerical procedures to estimate  
percentiles of sums of iidrv's
random variables: the Panjer recursion algorithm,
a method based on the Fast Fourier Transform,  
and Monte Carlo simulation \cite{klugman++_2004_loss,panjer_2006_operational,dupire1998monte}. 
These numerical techniques are efficient and yield accurate estimates
of high percentiles of sums of random variables provided 
that these are not too heavy-tailed: their computational cost  increases as the tails 
of the probability distribution become heavier, 
and eventually become impracticable.
When Monte Carlo simulation is 
used, this difficulty can be addressed using variance reduction techniques \cite{asmussen++_2000_rare,asmussen+kroese_2006_improved}. 

In this work we take a different approach and derive 
closed-form approximations for high percentiles
of the aggregate distribution based on a perturbative expansion. 
The zeroth order term in the perturbative expansion is similar to 
the single-loss approximation \cite{boecker+klueppeberg_2005_operational}, 
which assumes that the sum is dominated by the maximum. This dominance
in the sum by the maximum is a property of subexponential distributions, 
a subclass of heavy-tailed distributions \cite{goldie+klueppelberg_1998_subexponential,foss+etal_subexponential}. 
These types of distributions appear in important areas of application, 
such as insurance and finance  \cite{embrechts++_1997_modelling},
hydrology \cite{reiss2007statistical},
queueing models \cite{asmussen2003applied,tsourti+panaretos_2004_extreme_telefraffic},
the characterization of the Internet \cite{crovella+taqqu+bestavros_1998_heavyTailed},
and other areas of application \cite{resnick_2007_heavy}.

The first order perturbative approximation, which includes
the zeroth order term plus a first order correction, is similar
to approximations that can be derived from 
the asymptotic tail behavior of sums of subexponential variables
\cite{omey+willekens_1986_second,omey+willekens_1987_second,grubel_1987_subordinated,sahay++_2007_operational,degen_2010_calculation,barbe+mccormick_2005_asymptotic,barbe++_2007_asymptotic,barbe+mccormick_2009_asymptotic,albrecher++_2010_higherOrder}. Assuming
that the mean of the individual random variables in the sum is finite,
these approximations are all similar to the mean-corrected single-loss formula,
which was proposed by \cite{boecker+sprittulla_2006_operational} 
using heuristic arguments. 
In this article we provide an explicit procedure to derive
higher order terms in the perturbative expansion, which 
provides a more accurate approximation to
high percentiles of sums of positive iidrv's. 

The perturbative series introduced in this article
differs in important aspects from previous approximations
proposed in the literature. In particular,
the terms in the perturbative series are expressed as a function
of the moments of the right-truncated distribution for the
individual rv's in the sum. These 
censored moments exist even when the moments of the 
original distribution (without truncation) diverge. 
Consequently, the same expression is valid for both the finite and 
infinite mean cases. For high percentiles, the perturbative expansion provides a sequence of approximations that, up to certain order, has increasing quality as more terms are included. Beyond that order the convergence of the series deteriorates.  

The article is organized as follows: section \ref{sec:perturbativeExpansion} 
presents the derivation of a perturbative expansion for the percentile of 
sums of two random variables. 
This expansion is then applied to the estimation of
high percentiles of sums of $N$ independent random variables 
in section \ref{sec:maximumExpansion}. 
The key idea is to treat separately the maximum and 
the remaining terms in the sum. Explicit formulas are derived 
when $N$, the number of terms in the sum, 
is either deterministic or stochastic. 
Section \ref{sec:relatedWork} reviews
the approximations for high percentiles of sums of iidrv's
that have been proposed in the literature. 

The accuracy of the perturbative series is illustrated
in section \ref{sec:empiricalEvaluation} by comparing
with exact results or with Monte Carlo estimates, 
if closed-form expressions are not available.
Finally, section \ref{sec:conclusions} summarizes the contributions of 
this work and discusses the perspectives for further research.

\section{Perturbative expansion for the percentiles of 
the sum of two random variables} \label{sec:perturbativeExpansion}
In this section we derive a perturbative expansion of 
the percentile of a sum of two random variables. 
The zeroth order term in the perturbative series 
is the percentile of one of the variables in the sum.
Higher order terms involve the moments 
of the second variable, conditioned to the first one having
a fixed value.
In the following section, these general expressions
are applied to the particular case sums of $N$ random 
variables
by identifying the first random variable with 
the maximum in the sum and the second one
with the remainder. 

Let $X$ and $Y$ be two rv's whose joint 
distribution function is $F_{X,Y}(x,y)$ (density $f_{X,Y}(x,y)$). 
Consider the random variable
\begin{equation}
\label{EXPANSION_DEF}
Z = X + \epsilon Y, 
\end{equation}  
whose probability distribution is  $F_Z(z)$ (density $f_Z(z)$). 
It is not possible to express this distribution in 
a closed form that does not involve a convolution, 
except in special cases \cite{nolan_2012_stable,nadarajah_2006_distribution}. 
Let $Q_0 = F_{X}^{-1}(\alpha)$ and 
$Q = F_Z^{-1}(\alpha)$ be the $\alpha$-percentiles of $X$ and $Z$, respectively.
The percentile of $Z$ at probability level $\alpha$ 
can be formally represented by a power series in $\epsilon$
\begin{equation} 
\label{EXPANSION}
Q = Q_0 + \delta Q = Q_0 + \sum_{k=1}^{\infty} \frac{1}{k!}  Q_k {\epsilon}^{k} \, .
\end{equation}
The approximation of order $K$ to $Q$ is the result of keeping
only the first $K+1$ terms in the series
\begin{equation} 
Q^{(K)} \equiv  Q_0 + \sum_{k=1}^{K}  Q_k {\epsilon}^{k}/k! \, .
\end{equation}
Explicit expressions for the zeroth and first coefficients in 
(\ref{EXPANSION}) have been derived in \cite{GOURIEROUX}, 
in the context of credit risk. Also in this context, \cite{WILDE} 
give an explicit expression for the derivatives $ d^nF_Z(z)/ d\epsilon^n$,
which are used in the perturbative expansion in $\epsilon$ for $F_Z(z)$, 
the CDF of the sum.  Our goal in this section is to derive a 
general expression for the terms in a perturbative
expansion of the percentile (i.e. the inverse function $F_Z^{-1}(\alpha)$). 

The starting point of the derivation is the identity
\begin{equation}
\begin{aligned}
\label{STARTING_POINT}
0 & = F_Z(Q)-F_X(Q_0) \\
& = \int_{-\infty}^{\infty} dy \int_{Q_0}^{Q - \epsilon y}dx  f_{X,Y}(x,y).
\end{aligned}
\end{equation}
For a sufficiently smooth $f(x)$, one can define the operators
\begin{equation}
\begin{aligned}
e^{t \partial_x} f(x) & \equiv \sum_{k=0}^\infty \frac{t^k}{k!} \frac{\partial^k}{\partial x^k} f(x) = f(x + t) \\
\partial_x^{-1} f(x) & \equiv \int_{-\infty}^{x}  du \, f(u), \\
\end{aligned}
\end{equation}
where $\partial_x \equiv \frac{\partial \,}{\partial x}$, and their composition
\begin{equation}
\left(e^{t \partial_x} -1 \right) \partial_x^{-1} 
f(x)  = \int_{x}^{x + t} f(u) du.
\end{equation}
In terms of these operators
\begin{equation}
\begin{aligned}
 &   \int_{Q_0}^{Q - \epsilon y} dx  f_{X,Y}(x,y) =
  \int_{Q_0}^{Q_0 + \delta Q - \epsilon y} dx  f_{X,Y}(x,y) \\ 
 & = \left( e^{(\delta Q - \epsilon y) \partial_x } -1\right)\partial_x^{-1} f_{X,Y}(x , y) \Big\vert_{x=Q_0}.
\end{aligned}
\end{equation}
Using this result, (\ref{STARTING_POINT}) can be expressed as
\begin{equation}
\label{BASIC_EQUATION} 
 0  = \int_{-\infty}^{\infty} dy  \left. \left( e^{(\delta Q - \epsilon y) \partial_x } -1\right)\partial_x^{-1} f_{X,Y}(x , y) \right\vert_{x=Q_0}.
\end{equation}
Expanding the exponential operator in a formal Taylor power series
and using the definition of the complete Bell polynomials 
(Appendix \ref{app:bell}, eq. (\ref{A_BELL2}))
this expression becomes
\begin{equation}
\begin{aligned}
0 &= \int_{-\infty}^{\infty} dy \sum_{k=1}^{\infty} \frac{\epsilon^k}{k!} \times \\
 &B_k\left(\left(Q_1-y\right) \partial_x,Q_2 \partial_x , \ldots , Q_k \partial_x \right) \partial_x^{-1} f_{X,Y}(x,y) \Big\vert_{x=Q_0} \, .
\end{aligned}
\end{equation}
Since this equality holds for all $\epsilon$, each coefficient in the 
sum must be zero separately. This yields the system of equations
\begin{equation}
\begin{aligned}
\label{EXPANSION_IN_BELL} 
&0 = \int_{-\infty}^{\infty} dy  \\
&B_k \left(\left(Q_1 -y \right) \partial_x,Q_2 \partial_x,\ldots,Q_k \partial_x \right) \partial_x^{-1} f_{X,Y}(x,y) \Big\vert_{x=Q_0}, \\
&\text{for } \quad k\geq 1.
\end{aligned}
\end{equation}
Explicit expressions for $Q_k$  can be derived in terms of  $C_k$, 
a centered version  of the Bell polynomials (Appendix \ref{app:bell}, eq. (\ref{A_CENTRALBELL}))
\begin{equation} 
\begin{aligned}
\label{MASTER_Q}
Q_1 &= \EX \left[Y|X=Q_0\right] \\
Q_k &= - \frac{1}{f_X(Q_0)}\bigg[ \\
 & \hspace*{-0.4cm} \sum_{i=1}^{k} \binom{k}{i} 
C_{k-i}(Q_2\partial_x,\ldots,Q_{k-i}\partial_x)\partial_x^{i-1}\left\lbrace f_X(x)\tilde{M}_{i}(x) \right\rbrace \\
& \hspace*{-0.4cm} +\sum_{i=2}^{k-2} \binom{k-1}{i-1} 
Q_{i} C_{k-i}(Q_2\partial_x,\ldots,Q_{k-i}\partial_x) f_X(x) \bigg]_{x=Q_0}, \\ 
& \text{for }\ k\geq 2,  \\
\end{aligned}
\end{equation}
where 
\begin{equation}
\label{RARE_Y_MOMENTS}
\begin{aligned}
\tilde{M}_i(x) &\equiv \EX[(Q_1-Y)^i \vert X=x] \\ 
&= \sum_{j=0}^i \binom{i}{j} (-1)^j Q_1^{i-j} M_{j}(x)
\\
M_j(x) &\equiv \EX[Y^j\vert X=x].
\end{aligned}
\end{equation}
These recursive formulas for the coefficients and for $C_k$ can be 
used to compute the approximation to the percentile $Q$ to any order in $\epsilon$.
However, the complexity of the explicit formulas for the 
coefficients increases with their order. 
The first four terms in the perturbative series are
\begin{align}
\label{COEFFICIENTS}
Q_0 =& F_X^{-1}(\alpha) \nonumber \\
Q_1 =& \EX \left[Y|X=Q_0\right] \nonumber \\
Q_2 =& -\frac{1}{f_X(Q_0)}  \partial_x \left\lbrace f_X(x) \tilde{M}_2(x) \right\rbrace_{x=Q_0} \nonumber \\ 
    =& -\frac{1}{f_X(Q_0)}  \partial_x \left\lbrace f_X(x) \VA[Y|X=x] \right\rbrace_{x=Q_0}  \nonumber \\
Q_3 =& -\frac{1}{f_X(Q_0)} \Big\lbrace \partial_x^2 \left( f_X(x) \tilde{M}_3(x) \right) + \nonumber \\
    & 3Q_2\partial_x \left( f_X(x) \tilde{M}_1(x) \right) \Big\rbrace_{x=Q_0} \nonumber \\
Q_4 =& -\frac{1}{f_X(Q_0)} \Big\lbrace \partial_x^3 \left( f_X(x) \tilde{M}_4(x) \right) + \nonumber\\
    & 6Q_2\partial_x^2 \left( f_X(x) \tilde{M}_2(x) \right) + 4Q_3\partial_x \left( f_X(x) \tilde{M}_1(x) \right) + \nonumber\\
    & 3Q_2^2\partial_x f_X(x)   \Big\rbrace_{x=Q_0}.
\end{align}
The term $Q_2$  can be expressed in terms of the conditional variance
 ($\VA[Y|X=x]$) instead of $\tilde{M}_2(x)$ because, for this particular 
term, $Q_1$ can be replaced by
 $\EX[Y|X=x]$. This substitution is not possible in general for higher order terms.

These general expressions for the terms in a perturbative expansion
of the percentiles of the sum of two random variables
will be applied in the following section to sums of $N$ independent random variables, 
where $N$ can be deterministic or stochastic.  

\section{Perturbative expansion around the percentile of the maximum}
\label{sec:maximumExpansion}
In this section (\ref{COEFFICIENTS}) is used to estimate
high percentiles of the sums of  independent random variables with 
heavy tails
\begin{equation}
\label{Z_EQ}
Z_N = \sum_{i=1}^N L_i,
\end{equation}
where  $\{L_i\}_{i=1}^N$  are positive iidrv's sampled 
from $F(l)$ (the corresponding density is $f(l)$). 
Let $G(z)$ be the probability distribution of the sum $Z_N$, and  $g(z)$ 
the corresponding density. The key idea is to partition
the sum into two contributions: 
the maximum and the sum of the remaining terms
\begin{eqnarray}
\label{MAX_SPLITTING}
Z_N(\epsilon) & = &  X_N +\epsilon Y_N  \nonumber \\
X_N & = & L_{[N]} \equiv \max{\left[\left\{L_i\right\}_{i=1}^N \right]} 
\nonumber \\
Y_N & = & \sum_{i=1}^{N-1} L_{[i]} 
\end{eqnarray}
where  $L_{[i]}$ is the $i$-th order statistic of  the sample
$\left\{L_i\right\}_{i=1}^N$ 
(i.e. $ L_{[1]} \le L_{[2]} \le \ldots \le L_{[N]}$).
The formal parameter $\epsilon$  is introduced to order the terms 
in the perturbative expansion. It is eventually set to one ($\epsilon = 1$), 
so  that $Z_N(1) = Z_N$. As shown in Appendix \ref{app:relatedWork}, the 
perturbative series truncated to first order
provides an estimate that is similar to 
approximations that can be derived from 
the tail behavior of sums of subexponential variables
\cite{sahay++_2007_operational,degen_2010_calculation,albrecher++_2010_higherOrder}.
Therefore, the analysis presented in \cite{omey+willekens_1986_second,omey+willekens_1987_second}
can be used to establish the asymptotic properties of this approximation.
The issue of convergence of the perturbative series 
outside of the asymptotic regime is analyzed empirically in 
section \ref{sec:empiricalEvaluation}.
Qualitatively, the perturbation term in (\ref{MAX_SPLITTING}) 
is small if $L_{[N]} \gg \sum_{i=1}^{N-1} L_{[i]} $;
that is, when the sum (\ref{Z_EQ}) is dominated by the maximum. 
This is the case when the probability distribution of $L$ is subexponential, 
provided that the value of the sum is sufficiently large 
\cite{goldie+klueppelberg_1998_subexponential,foss+etal_subexponential}. 
In consequence, the perturbative series should be more accurate for high percentiles. The empirical analysis carried out reveals that, for sufficiently high percentiles, the accuracy of the approximation initially improves as more terms are included in the series. However, beyond a certain  order the approximation actually becomes worse when further terms are used, which indicates that, in the cases studied, the perturbative series is not convergent. 

The probability distribution of the maximum $L_{[N]}$ is  
\begin{equation} \label{eq:maxPDF}
F_{[N]}(x) = F(x)^N.
\end{equation}
The corresponding density is obtained by taking 
the derivative of (\ref{eq:maxPDF})
\begin{equation}
f_{[N]}(x) = N {F(x)}^{N-1}f(x).
\end{equation}
In terms of these, the perturbative expansion 
 (\ref{MASTER_Q}) becomes
\begin{equation} 
\begin{aligned} \label{eq:mainResult}
Q_0 & = F^{-1}(\alpha^{\frac{1}{N}}) \\
Q_1 &= \EX \left[\sum_{i=1}^{N-1} L_{[i]}|L_{[N]}=Q_0\right] \\
Q_k &= -\frac{1}{f_{[N]}(Q_0)}\bigg[ \\
& \hspace*{-0.4cm} \sum_{i=1}^{k}
\binom{k}{i} C_{k-i}(Q_2\partial_x,\ldots,Q_{k-i}\partial_x)
\partial_x^{i-1}\left\lbrace f_{[N]}(x)\widetilde{M}_{i}(x) \right\rbrace \\
&\hspace*{-0.4cm} +\sum_{i=2}^{k-2} \binom{k-1}{i-1} Q_{i} C_{k-i}(Q_2\partial_x,\ldots,Q_{k-i}\partial_x) f_{[N]}(x) \bigg]_{x=Q_0},\\
& \text{for }\ k\geq 2,  
\end{aligned}
\end{equation}
with 
\begin{equation}
\begin{aligned}
\widetilde{M}_i(x) &\equiv \EX\Big[ \Big( Q_1-\sum_{k=1}^{N-1} L_{[k]} \Big)^i \Big\vert  L_{[N]}=x \Big]\\  
  &= \sum_{j=0}^i \binom{i}{j} (-1)^j Q_1^{i-j} M_{j}(x),
\end{aligned}
\end{equation}
where $ M_j(x) $ is the $j$th conditional
moments of the random variable $Y_N = \sum_{i=1}^{N-1} L_{[i]}$
\begin{equation} \label{eq:lowerStatisticsmoments}
M_j(x)  \equiv \EX\left[\left. \left(  \sum_{i=1}^{N-1} L_{[i]} \right)^j \right| 
L_{[N]} =x \right].
\end{equation}
These closed-form expressions for the terms in the perturbative series 
(\ref{eq:mainResult}) are the main contribution of this research. 
Explicit formulas
for the conditional moments (\ref{eq:lowerStatisticsmoments})
can be readily obtained using the invariance of 
$ \sum_{i=1}^{N-1} L_{[i]} $ under
an arbitrary permutation of the indices  
\begin{equation}
\begin{aligned}
\label{CENSORED}
M_j(x)  & = \EX \Big[ \Big(  \sum_{i=1}^{N-1} L_{[i]} \Big)^j \Big\vert 
L_{[N]} = x \Big] \\
        &= \EX\Big[ \Big( \sum_{i=1}^{N-1}L_{i} \Big)^j \Big\vert \big\lbrace L_{i} \le x \big\rbrace_{i=1}^{N-1}\Big] \\
& =   \int_0^x d l_1 \ldots \int_0^x d l_{N-1} \ 
 \left(  \sum_{i=1}^{N-1}l_{i} \right)^j \prod_{i=1}^{N-1}  \frac{f(l_i)}{F(x)}.
\end{aligned}
\end{equation}
The last quadrature is the average of
the $j$th power of the sum of $N-1$ independent random variables  
$ \left\{ L_{i} \right\}_{i=1}^{N-1} $,
whose joint distribution is 
\begin{equation}
\begin{aligned}
\label{PROPOSITION1}
f \Big( \big\lbrace l_{i} \big\rbrace_{i=1}^{N-1} \Big\vert 
\big\lbrace l_{i} \le x \big\rbrace_{i=1}^{N-1} \Big) 
&= \prod_{i=1}^{N-1}f(l_i|l_i \leq x) \\
&= \prod_{i=1}^{N-1} \frac{f(l_i)}{F(x)} \theta(x-l_i),
\end{aligned}
\end{equation}   
where $ \prod_{i=1}^{N-1} \theta(x-l_i) $ is 
a product of Heaviside step functions, 
which is equal to $1$ in the region $\left\{ l_i \leq x \right\}_{i=1}^{N-1}$ 
and $0$ outside this region. 
Using the definition of the complete Bell polynomials 
(\ref{A_MOMENTSBELL}),
it is possible to express the $j$th moment of 
the sum $\sum_{i=1}^{N-1}l_{i} $,
where the terms in the sum
are constrained to be in the region $\left\{ l_i \leq x \right\}_{i=1}^{N-1}$,
\begin{align}
M_j(x) &= B_j \left( K_{1}(x),\ldots,K_{j}(x) \right), 
\end{align}
in terms of the conditional cumulants $K_j(x)$, defined as
\begin{equation}
K_j(x)  = \left. \frac{d^j \,}{ds^j} \left[\log \left(\int_{0}^{\infty} dy  \, e^{s y} f_{Y_N \vert X_N}(y \vert x) \right) \right] \right|_{s = 0}.
\end{equation} 
Finally, using the property that the $p$th cumulant of a sum of 
independent variables is the
sum of the $p$th cumulants of the individual variables 
\begin{equation}
K_{p}(x) = (N-1) \kappa_p(x),  \quad p = 1,2,\ldots
\end{equation}
we obtain
\begin{align}
M_j(x) &= B_j \left( (N-1)\kappa_1(x),\ldots,(N-1)\kappa_j(x) \right),
\end{align}
where $\kappa_j(x)$ is the $j$th censored cumulant of $L$ 
\begin{equation} \label{eq:censored_cumulant}
 \kappa_j(x)  = \left. \frac{d^j \,}{ds^j} \left[\log \left(\int_{0}^{x} dl  \, e^{s l} \frac{f(l)}{F(x)}  \right) \right] \right|_{s = 0}.
\end{equation}
These censored cumulants can also be expressed in terms of the 
censored moments of $L$
\begin{eqnarray}
 \kappa_j(x)     & = & \mu_j(x) - \sum_{i=1}^{j-1} \binom{j-1}{i}  \kappa_{j-i}(x) \mu_i(x),  \\
 \mu_j(x) & = & \int_{0}^{x} dl  \, l^j \frac{f(l)}{F(x)} , \quad \text{for} \quad  j = 1,2, \ldots
\end{eqnarray}

Using these relations, it is possible to derive explicit formulas 
for the terms in the perturbative series. 
In particular, the first three are
\begin{eqnarray}
Q_0 & = & F^{-1}(\alpha^{\frac{1}{N}}) \label{eq:Q0}\\
Q_1 & = & (N-1)\EX \left[L|L \leq Q_0\right] \label{eq:Q1}\\
Q_2 & = & - \frac{N-1}{{F(Q_0)}^{N-1}f(Q_0)}  \nonumber\\
& & \partial_x \left( {F(x)}^{N-1}f(x) \VA \left[L|L \leq x\right] \right)_{x=Q_0}  \nonumber \\
& = & - \left(N-1\right)  \Big[ \nonumber\\
 & & \left((N-2) \frac{f(Q_0)}{F(Q_0)}  + \frac{f'(Q_0)}{f(Q_0)}  \right) 
 \VA \left[L|L \leq Q_0\right] + \nonumber \\
&& \frac{f(Q_0)}{F(Q_0)} \left(Q_0- \EX \left[L|L \leq Q_0\right]\right)^2 
 \Big].  
\end{eqnarray}
An attractive feature of this expansion is that the approximation of order $K$ 
depends only on the censored moments of $F$ of order lower or equal to $K$.  
Since they are censored, these always exist, even for distributions 
whose moments diverge.
These expressions have been obtained for cases in which the number 
of terms in the sum (\ref{Z_EQ}) is fixed. 
In the next section, we derive closed-form 
expressions for sums with a random number of terms.

\subsection{Sums with a random number of terms}
In many applications the quantities of 
interest are aggregate random variables consisting
of a variable number of terms 
\begin{equation} \label{eq:sum_randomFrequency}
Z_N = \sum_{i=1}^N L_i,
\end{equation}
where $N$ is a discrete random variable 
whose probability mass function is
\begin{equation} 
P[N=n] \equiv p_n \ , \ \ n=0,\ldots,\infty.
\end{equation}
In insurance and operational risk 
\cite{embrechts++_1997_modelling,mcneil2005},
where $Z_N$ represents the aggregate loss in a fixed time period
(e.g. yearly losses), $N$ is referred to as the \textit{frequency} 
of the loss events. For convenience, we will use this
term to refer to $N$ in the remainder of the article.

Consider the random variable $Z_N=X_N+Y_N$, with
\begin{eqnarray}
X_N & = & L_{[N]}\\
Y_N & = & \sum_{i=1}^{N-1} L_{[i]},
\end{eqnarray}
as in (\ref{Z_EQ},\ref{MAX_SPLITTING}), where $N$ is now a 
integer random variable. 
We denote $X_n = L_{[n]}$ and $Y_n = \sum_{i=1}^{n-1} L_{[i]}$ 
the corresponding random variables conditional on a fixed value $N=n$. 
In terms of the probability distribution of 
$L_{[n]}$, the probability distribution of the maximum of the $n$ terms 
in the sum ($F_{[n]}(x) =F(x)^n$), and of the corresponding density
($f_{[n]}(x) = n F(x)^{n-1} f(x)$),
the probability distribution and the density of $L_{[N]}$ are
\begin{equation}
F_{[N]}(x) = \sum_{n=0}^{\infty} p_n F_{[n]}(x) \quad 
f_{[N]}(x) = \sum_{n=0}^{\infty} p_n f_{[n]}(x),
\end{equation} 
respectively.

For random $N$ the zeroth order term in the perturbative expansion 
$Q_0$ satisfies the relation
\begin{equation}
\begin{aligned}
\label{eq:Q0_randomFrequency_implicit}
\alpha &= F_{[N]}(Q_0) = \sum_{n=0}^{\infty} p_n F_{[n]}(Q_0)= \sum_{n=0}^{\infty} p_n F(Q_0) ^n   \\
&= \EX \left[ F(Q_0)^{N} \right]= \mathcal{M}(\log{F(Q_0)}),
\end{aligned}
\end{equation}
where $\mathcal{M}_N(s)$ is the moment generating function of the random variable $N$
\begin{equation}
\mathcal{M}_N(s) \equiv \mathtt{E}\left[ e^{sN} \right]  = \sum_{n=0}^{\infty} p_n e^{sn}.
\end{equation}

Using this definition we can invert (\ref{eq:Q0_randomFrequency_implicit})
\begin{equation}  \label{eq:Q0_mgf}
Q_0 = F^{-1}(e^{\mathcal{M}_N^{-1}(\alpha)}).
\end{equation}
Starting from (\ref{EXPANSION_IN_BELL}) with $k=1$ it is possible derive an expression
for the first term in the perturbative series in terms of $Q_0$
\begin{equation} 
\begin{aligned}
&Q_1\sum_{n=0}^{\infty} p_n f_{[n]}(Q_0) =\\  &\sum_{n=0}^{\infty} p_n f_{[n]}(Q_0) \EX \Big[\sum_{i=1}^{n-1} L_{[i]} \Big\vert L_{[n]} = Q_0 \Big].
\end{aligned}
\end{equation}
Using the explicit form of the probability 
distribution of the maximum and equation (\ref{CENSORED}), we get  
\begin{equation}
Q_1 = \frac{\EX \left[ N (N-1) F^N(Q_0) \right]}{\EX \left[  N F^N(Q_0) \right] } \EX[L|L \leq Q_0].
\end{equation}
For the higher order coefficients an analogous derivation
from (\ref{MASTER_Q}) yields 
\begin{eqnarray}
-Q_k \sum_{n=0}^{\infty} p_n f_{[n]}(Q_0) = \left[ 
\sum_{s=1}^{k} \binom{k}{s} C_{k-s}(\ldots) \partial_x^{s-1} U_s(x) \right. \nonumber \\
\left. +  \sum_{s=2}^{k-2} \binom{k-1}{s-1} Q_{s} C_{k-s}(\ldots)  \sum_{n=0}^{\infty} p_n f_{[n]}(x)  \right]_{x=Q_0}
\end{eqnarray}
where 
\begin{equation} 
\begin{aligned}
U_s(x) =& \sum_{n=0}^{\infty} p_n f_{[n]}(x)\EX[(Q_1-Y_n)^s |X_n=x] \\
       =& \sum_{n=0}^{\infty} p_n f_{[n]}(x)\sum_{q=0}^s \binom{s}{q} (-1)^q Q_1^{s-q}M_{n,q}(x) \\
M_{n,q}(x) \equiv & B_q \left( (n-1)\kappa_1(x),\ldots,(n-1)\kappa_q(x) \right).
\end{aligned}
\end{equation}
To compute the expected values over the frequency, one needs
to isolate the dependency on $N$. For this purpose, it is
convenient to use an alternative representation of the Bell polynomials
that allows to express moments
in terms of cumulants using partitions of sets
(Appendix \ref{app:bell}, eq. (\ref{A_PARTITIONSBELL}))
\begin{equation}
 M_{n,q}(x) = \sum_{\Pt \in \PtsS(q)}(n-1)^{|\Pt|} \prod_{\elemPt \in \Pt}\kappa_{|\elemPt|}(x),
\end{equation}
where $\PtsS(q)$ is the set of all partitions of the set ${1,2,\ldots,q}$, and $|\mathtt{A}|$ and $|\mathtt{b}|$ denote the number of elements in 
the sets $\mathtt{A}$ and $\mathtt{b}$ respectively, and 
$\kappa_{|\elemPt|}(x)$ is the $|\elemPt|$th  censored cumulant
of $L$, as defined in (\ref{eq:censored_cumulant}).

Using this expression the coefficients become
\begin{equation} 
\begin{aligned}
\label{MASTER_STOCHASTIC}
Q_1 &= \frac{\lambda_1(Q_0)}{\lambda_0(Q_0)}\kappa_1(Q_0) \\
Q_k  &= \frac{-1}{\lambda_0(Q_0)}\Bigg[ \\
& \sum_{s=1}^{k} \binom{k}{s} C_{k-s}(\ldots) \partial_x^{s-1}  \sum_{q=0}^s  
\binom{s}{q}(-1)^q Q_1^{s-q} \times\\
&\sum_{\Pt \in \PtsS(q)} \lambda_{|\Pt|}(x) \prod_{\elemPt \in \Pt}  \kappa_{|\elemPt|}(x) \\
&  +  \sum_{s=2}^{k-2} \binom{k-1}{s-1} Q_{s} C_{k-s}(\ldots)  \lambda_0(x)  \Bigg]_{x=Q_0}, \ \ \text{for }\ k\geq 2, \\
\end{aligned}
\end{equation}
with 
\begin{equation} 
\begin{aligned}
\label{LAMBDA_DEFINITION}
\lambda_a(x) &\equiv \EX_N [(N-1)^{a}f_{[N]}(x) ] \\
&= \frac{f(x)}{F(x)} \EX [N(N-1)^a{F(x)}^N ],\ \ \text{for }\ a\geq 0.
\end{aligned}
\end{equation}
The explicit expressions for the first four coefficients are
\begin{equation} 
\begin{aligned} \label{eq:Q0_4_randomFrequency}
Q_0 & = F^{-1}(e^{\mathcal{M}_N^{-1}(\alpha)}) \\
Q_1 &= \frac{\lambda_1(Q_0)}{\lambda_0(Q_0)}\kappa_1(Q_0) \\
Q_2  &= \frac{-1}{\lambda_0(Q_0)}\partial_x \Big[ Q_1^2 \lambda_0 -2Q_1 \lambda_1 \kappa_1 + \lambda_1\kappa_2 + \lambda_2 \kappa_1^2 \Big]_{x=Q_0} \\
     &= \frac{-1}{\lambda_0(Q_0)} \partial_x\left[ \lambda_1 \kappa_2 +(\lambda_2 - \frac{\lambda_1^2}{\lambda_0})\kappa_1^2\right]_{x=Q_0}\\
Q_3  &= \frac{-1}{\lambda_0(Q_0)} \Bigg\{ 3Q_2\partial_x \Big[ Q_1\lambda_0 -\lambda_1\kappa_1 \Big]_{x=Q_0} \\
&+\partial_x^2 \Big[ Q_1^3\lambda_0-3Q_1^2\lambda_1\kappa_1 + 3Q_1(\lambda_1 \kappa_2 + \lambda_2{\kappa_1}^2 ) \\
&- \lambda_1\kappa_3 - 3\lambda_2 \kappa_1\kappa_2 - \lambda_3{\kappa_1}^3 \Big]_{x=Q_0}\Bigg\},
\end{aligned}
\end{equation}
where, to simplify the notation, the  dependence  on $x$ in the 
$\lambda_a(x)$ and $\kappa_b(x)$ has been omitted.

The functions $\left\{\lambda_a(x); \ a = 0,1,2,\ldots \right\}$ 
can also be expressed in terms of the moment generating function of $N$ as  
\begin{equation}
\label{LAMBDA_FROM_MGF}
\lambda_a(x)  =  \frac{f(x)}{F(x)} \partial_s \left. \left( \partial_s-1 \right)^a \mathcal{M}_N(s) \right\vert_{s=\log{F(x)}} \quad \text{for }\ a\geq 0.
\end{equation}
Explicit expressions for the Poisson and negative 
binomial probability distributions are given in Appendix \ref{app:frequency}. 
These types of  distributions are commonly used in applications.

\subsection{Approximation in terms of frequency moments for high percentiles} 

The formulas derived in the previous section (\ref{eq:Q0_4_randomFrequency})
are different from the standard single-loss approximation
\cite{boecker+klueppeberg_2005_operational}
and corrections thereof 
\cite{boecker+sprittulla_2006_operational,sahay++_2007_operational,degen_2010_calculation,albrecher++_2010_higherOrder}. In this section we show that for high percentiles
one recovers the single-loss approximation and correction
terms. In the limit $\alpha \rightarrow 1^{-}$ 
the inverse of the moment generating function
in (\ref{eq:Q0_mgf}) can be approximated as 
\begin{equation}
\mathcal{M}_N(s) = \mathtt{E}\left[ e^{sN} \right]  
= 1 + s \mathtt{E}\left[ N \right] + {\cal O}(s^2), 
\end{equation}
for $s \rightarrow 0$.
From this expression,
\begin{equation}
\mathcal{M}_N^{-1}(\alpha) \approx -\frac{1-\alpha}{\EX [N]},
\quad \text{for} \quad  \alpha \rightarrow 1^-.
\end{equation}
This leads to the standard single-loss approximation
\cite{boecker+klueppeberg_2005_operational}
\begin{equation}
Q_0 \approx Q_{SL} \equiv F^{-1} \left(1 -\frac{1-\alpha}{\EX [N]} ) \right).
\end{equation}

In this limit, the survival
function  $ S(x) \equiv 1-F(x) $ approaches $ 0$, and simpler approximate expressions for 
$\lambda_a(x)$ are obtained by keeping terms only up to 1st order in  $S(x)$ 
\begin{equation} 
\begin{aligned}
\lambda_a(x) &= \partial_x \EX [(N-1)^a{(1-S(x))}^N] \\
 &\approx \partial_x \EX [N(N-1)^a(1-NS(x)] \\
 &=-\partial_x S(x) \EX[N(N-1)^a] \\
 &= f(x) \sum_{s=0}^a \binom{a}{s} (-1)^{a-s}\nu_{s+1}, \ a = 0,1,\ldots
\end{aligned}
\end{equation}
where $\nu_s=\EX[N^s]$ are the moments of the frequency distribution. 
Using these approximations, the high-percentile corrections to 
the single-loss formula can be expressed directly in terms
of the moments of the frequency distribution 
\begin{eqnarray}
Q_1 &\approx & \left(\frac{\EX[N^2]}{\EX[N]} - 1 \right) \EX[L|L \leq Q_0] 
\label{eq:Q1_highpercentile}\\
Q_2 &\approx & - \left(\frac{\EX[N^2]}{\EX[N]} -1\right)
\frac{1}{f(x)}\partial_x\left[ f(x) 
\VA \left[L|L \leq x\right] \right]_{x=Q_0} \nonumber \\
& & +
\left( \left(\frac{\EX[N^2]}{\EX[N]} \right)^2 - 
\frac{\EX[N^3]}{\EX[N]} \right)\times \nonumber\\
& &  \hspace*{0.3cm}\frac{1}{f(x)}\partial_x\left[ f(x) \EX[L|L \leq x]^2\right]_{x=Q_0}.
\end{eqnarray}
The approximation to $Q_1$ is similar to the corrections to the single
loss formula proposed in the literature
\cite{boecker+sprittulla_2006_operational,sahay++_2007_operational,degen_2010_calculation,albrecher++_2010_higherOrder}. In section \ref{sec:relatedWork}, we provide
a review of these corrections. Their accuracy will be compared to the
perturbative expansion in section \ref{sec:empiricalEvaluation}. 
To make the numerical computation of the perturbative approximation  
up to high orders feasible it is useful to 
express the terms of the series recursively. 
These recursive expressions are presented in Appendix \ref{RECURSIVE_FORMULAE}.

\section{Related work} \label{sec:relatedWork}

In this section we review closed-form approximations
for the percentile of sums of positive iidrvâ's that have
been proposed in previous investigations. 
Even though it is possible to derive approximations 
for particular heavy-tailed distributions, such as \cite{Blum_1970}
for the Pareto distribution,
in this work we consider comparisons only with
approximations for general subexponential distributions
\cite{goldie+klueppelberg_1998_subexponential,foss+etal_subexponential}.
The single-loss approximation 
can be derived
using first order asymptotics of the tail of sums 
of subexponential random variables \cite{chistyakov_1964_theorem,embrechts+veraverbeke_1982_estimates,boecker+klueppeberg_2005_operational}.
Higher order asymptotic expansions of the tails of the 
compound distribution \cite{omey+willekens_1986_second,omey+willekens_1987_second,grubel_1987_subordinated,barbe+mccormick_2005_asymptotic,barbe++_2007_asymptotic,barbe+mccormick_2009_asymptotic}
can be used to obtain corrections
to the single-loss approximation
\cite{sahay++_2007_operational,degen_2010_calculation,albrecher++_2010_higherOrder}. 
These high order corrections are similar to the successive terms 
in the perturbative expansion analyzed in this article. 
However, there are some important differences. 
In particular, these terms are expressed as a function of right-censored moments,
which are always finite.  In  the section on experimental evaluation (section \ref{sec:empiricalEvaluation}) we will further show that the perturbative series
provides more accurate approximations than the expressions introduced in this section.

One of the defining properties of subexponential distributions
is that large values of sums of  subexponential random variables
are dominated by the maximum 
\begin{equation}
Z_N = \sum_{i=1}^N L_i \approx \max\left\{L_1,\ldots L_N\right\}, \quad Z_N \rightarrow \infty. 
\end{equation}
In insurance mathematics this  corresponds to the 'one loss causes ruin' 
regime \cite{embrechts++_1997_modelling}. Using the property of 
subexponential distributions 
\cite{chistyakov_1964_theorem,embrechts+veraverbeke_1982_estimates}
\begin{equation}
\lim_{x\rightarrow \infty} \frac{P(L_1+\ldots + L_N > x)}{P(L_1 > x)} = N,
\end{equation} 
it is possible to show that, for this type of distributions, the percentile 
of $Z_N$ at the probability level $\alpha$ 
is approximately 
\begin{equation} \label{eq:singleLoss_deterministic}
Q_{SL}  = F^{-1}\left(1 - \frac{1-\alpha}{N} \right), \quad \text{for}  \ \alpha \rightarrow 1^{-}.
\end{equation} 
In this limit,  expression (\ref{eq:singleLoss_deterministic})
is very similar to the zeroth order term in the perturbative expansion 
\begin{equation}
\begin{aligned}
Q_0 &= F^{-1}\left( \alpha^{\frac{1}{N}} \right) \\
&= F^{-1}\left( 1 - \frac{1-\alpha}{N}  + {\cal O} \left(\frac{(1-\alpha)^2}{N} \right) \right) \approx  Q_{SL}.
\end{aligned}
\end{equation} 
The derivation of a closed-form approximation for high percentiles 
using first order tail asymptotics can be readily extended 
to sums of subexponential iirdv's with a random number of terms
\begin{equation} \label{eq:singleLoss}
Q_{SL} = F^{-1}\left(1 - \frac{1-\alpha}{\mathtt{E}\left[N\right]} \right),
\end{equation} 
where $\mathtt{E}\left[N\right]$ is the average number of terms in the sum.
In the area of operational risk, this expression is known as the 'single-loss approximation' 
 \cite{boecker+klueppeberg_2005_operational,boecker+sprittulla_2006_operational}. 

Using heuristic arguments, a correction to the single-loss approximation
was proposed in 
\cite{boecker+sprittulla_2006_operational} for distributions with finite mean
\begin{equation} \label{eq:singleLossMeanCorrected}
Q \approx F^{-1}\left(1 - \frac{1-\alpha}{\mathtt{E}\left[N\right]} \right) 
+ \left(\mathtt{E}\left[N\right] -1 \right) \mu_L,  \quad \mu_L \equiv \mathtt{E}\left[ L \right].
\end{equation} 
In the limit  $\alpha \rightarrow 1^{-}$, the value $Q_0$ is large, so that
$\mathtt{E}\left[L | L \le Q_0 \right] \approx \mathtt{E}\left[L \right]$ and
the approximation given by (\ref{eq:singleLossMeanCorrected})
becomes similar to (\ref{eq:Q1_highpercentile}). 

Besides the heuristic derivation given in 
\cite{boecker+sprittulla_2006_operational}
and the perturbative expansion proposed in this work, 
higher order corrections to the single-loss approximation 
can be derived in at least three
different ways: Using the second order asymptotic
approximations introduced in
\cite{omey+willekens_1986_second,omey+willekens_1987_second,sahay++_2007_operational,degen_2010_calculation}, 
from the asymptotic expansion analyzed in 
\cite{barbe+mccormick_2005_asymptotic,barbe++_2007_asymptotic,barbe+mccormick_2009_asymptotic} 
or from asymptotic approximations based on evaluations
of $F(l)$ at different arguments \cite{albrecher++_2010_higherOrder}. 

In the case of distributions with finite mean, 
the asymptotic analysis of the tail of a 
subordinated distribution analyzed in
\cite{omey+willekens_1987_second} 
can be used to obtain $ Q_{OW}$,
a second order approximation of the percentile of sums 
of subexponential iidrv's, as the solution of
\begin{equation} \label{eq:OW_finiteMean} 
Q_{OW} = 
F^{-1}\left[1 - \frac{1-\alpha}{\mathtt{E}\left[N\right]} + 
\left(\frac{\mathtt{E}\left[N^2 \right]}{\mathtt{E}\left[N \right]} -1 \right)  
\mu_L f \left(Q_{OW} \right) \right]. 
\end{equation} 
This implicit nonlinear equation can be  solved numerically using, 
for example, an iterative scheme. 
Alternatively, one can retain only the leading terms
in a perturbative expansion of this expression
\begin{eqnarray} \label{eq:OW_finiteMean_perturbative}
Q_{OW}^{*}  & = & 
Q_{SL} 
+  
\left(\mathtt{E}\left[N \right] + (D-1) \right) \mu_L,
\end{eqnarray}
where $ D = \mathtt{Var}\left[N \right]/\mathtt{E}\left[N \right]$ 
is the index of dispersion
($D = 1$ for the Poisson distribution and $ D > 1$ for the negative binomial distribution).
The first term in (\ref{eq:OW_finiteMean_perturbative}) is the single-loss 
approximation \cite{boecker+klueppeberg_2005_operational,boecker+sprittulla_2006_operational}. 
The second term is a correction that involves the mean
and is similar to (\ref{eq:singleLossMeanCorrected})  
when $\mathtt{E}\left[N \right] \gg 1$ and $D \approx 1$. 
As shown in Appendix \ref{app:relatedWork},
expression (\ref{eq:OW_finiteMean_perturbative})
can be derived in a number of different ways
\cite{degen_2010_calculation,barbe+mccormick_2005_asymptotic,barbe++_2007_asymptotic,barbe+mccormick_2009_asymptotic, albrecher++_2010_higherOrder}. 

In the case of distributions with infinite mean,
in which the density is regularly varying at infinity with index 
 $-(1+a) $,  $ f(L) \in RV_{-(1+a)}$ \cite{Bingham1987}, the second order approximation of $Q \equiv G^{-1}(\alpha)$ satisfies the relation \cite{omey+willekens_1986_second}
\begin{equation} \label{eq:OW_infiniteMean}
\begin{aligned}
Q_{OW}  &=  F^{-1}\Bigg( 1- \frac{1- \alpha}{\mathtt{E}\left[N \right]}  \\
& \left. + c_a \left( \frac{\mathtt{E}\left[N^2\right]}{\mathtt{E}\left[N \right]}  -1 \right) \mu_F(Q_{OW}) f(Q_{OW}) \right), 
\end{aligned}
\end{equation}
where
\begin{equation}
\mu_{F}(x) \equiv \int_0^x ds (1-F(s)) = 
(1-F(x)) x + F(x) \mathtt{E}\left[L | \le x \right],
\end{equation}
and 
\begin{equation}
c_a = \left\{ 
\begin{array}{ll}
 1 & a = 1 \\
(1- 1/a) {\frac{\left[\Gamma(1-a) \right]^2}{2 \Gamma(1-2a)}} 
& a < 1
\end{array}
\right. ,
\end{equation}
where $\Gamma(x)$ is the gamma function.
Besides numerical schemes, an approximate closed-form expression
of the percentile, $Q_{OW}^{*}$, can be obtained using a 
perturbative scheme analogous to the 
finite mean case
\begin{eqnarray} \label{eq:OW_infiniteMean_perturbative}
Q_{OW}^{*}  & = & Q_{SL} 
+ c_{a} 
\left(\mathtt{E}\left[N \right] + (D-1) \right)
\mu_F(Q_{SL}), \\
\mu_F(Q_{SL})  & = &  
\frac{1-\alpha}{\mathtt{E}\left[N\right]} Q_{SL}  + 
\left(1 - \frac{1-\alpha}{\mathtt{E}\left[N\right]} \right) \mathtt{E}\left[L | L \le Q_{SL} \right].\nonumber\\
\end{eqnarray}
Appendix \ref{app:relatedWork} presents the detailed derivations of 
these approximations and the connections with the perturbative 
approach introduced in the current article. The main difference with 
previous proposals is that the perturbative expansion involves 
the moments of right-\linebreak truncated distributions. 
Since these censored moments are always finite, the same expressions
are valid for distributions with finite and with infinite mean. 
As illustrated in the following section, the
perturbative expansion provides accurate 
approximations of high percentiles 
of sums of iidrv's for a variety distributions and
a wide range of parameters, regardless of whether the mean
of the random variables in the sum is finite or infinite.

\section{Empirical evaluation} \label{sec:empiricalEvaluation}
In this section we investigate the properties of the perturbative
expansion of the $ \alpha $-percentile of the aggregate 
distribution introduced in this work, when 
$\alpha$ is close to $1$. The accuracy of 
this perturbative expansion is compared to the second order
asymptotic approximations
(\ref{eq:OW_finiteMean}-\ref{eq:OW_infiniteMean_perturbative})
for different types of distributions and
different values of $\alpha$. 
The types of distributions, ranges of parameters 
and percentile levels used to carry out the empirical 
evaluation of the proposed approximations are in the
range of those commonly used in applications in insurance 
and finance \cite{embrechts++_1997_modelling,mcneil2005}, 
especially in the area of operational risk
\cite{frachot++_2001_loss,embrechts++_2003_quantifying,panjer_2006_operational,carrilloMenendez+suarez_2012_robust}.
The derivation closed-form approximations 
for the estimation of high percentiles in these areas
of application is extremely relevant because of the 
large computational costs of the standard methods, such as MC 
simulation, which are used to compute the risk measures.

The comparisons among the different approximations 
are made in terms of the relative error $ (Q_{approx} - Q) / Q $,
where $Q_{approx}$ is an approximation of the percentile (either
$Q_{OW}$ $Q_{OW}^{*}$ or $Q^{(K)}$, the truncation of the perturbative 
series at order $K$), and $Q$ is the exact percentile. 
The sign of the error is retained in most cases 
to make it clear whether the approximation over- or underestimates  
the true value of the percentile.
When the true value of the percentile cannot be computed exactly,
it is estimated via Monte Carlo simulation. Due to the heavy-tailedness
of the severity distributions considered, many 
simulations are required to achieve sufficient 
precision in the percentile estimation. The Monte
Carlo estimates have been obtained using  
OpVision\textregistered\footnote{ www.opvision.es}, 
a software system for the 
analysis and quantification of operational risk
in the Advanced Measurement Approaches (AMA) 
framework \cite{basel2_2006_international}.
In all cases, the error of the Monte Carlo estimates 
is at most 0.1\% at a 95\% confidence level.
If the approximations analyzed are more accurate
than this threshold, more simulations are performed
to obtain reliable estimates of the accuracy. 
Error bands for the Monte Carlo estimates are displayed in all 
the graphs except for the L\'evy case, where the 
percentiles can be calculated exactly.
In many cases these sampling errors are much smaller
than the errors of the approximations considered 
and this band cannot be discerned in the plots.

The recursive formulas used for the calculation 
of the terms in the perturbative expansion are given in
Appendix \ref{RECURSIVE_FORMULAE}.
The computational cost of obtaining an approximation
with $K$ terms is $\mathcal{O}(K^4)$, where $K$ is the order 
at which the perturbative series is trunctated. 
An implementation in MatLab of the perturbative 
expansion is publicly available 
\footnote{www.qrr.es/technical-reports/QRR-2012-0001/code/perturbativeExpansion.m}.
In the experiments reported, the computations are 
numerically stable. However, numerical instabilities 
eventually appear for higher orders, higher quantiles 
and/or heavier-tailed distributions.

\begin{figure}[!ht]
\begin{center}
\begin{tabular}{cc}
\includegraphics[width=84mm,clip=true]{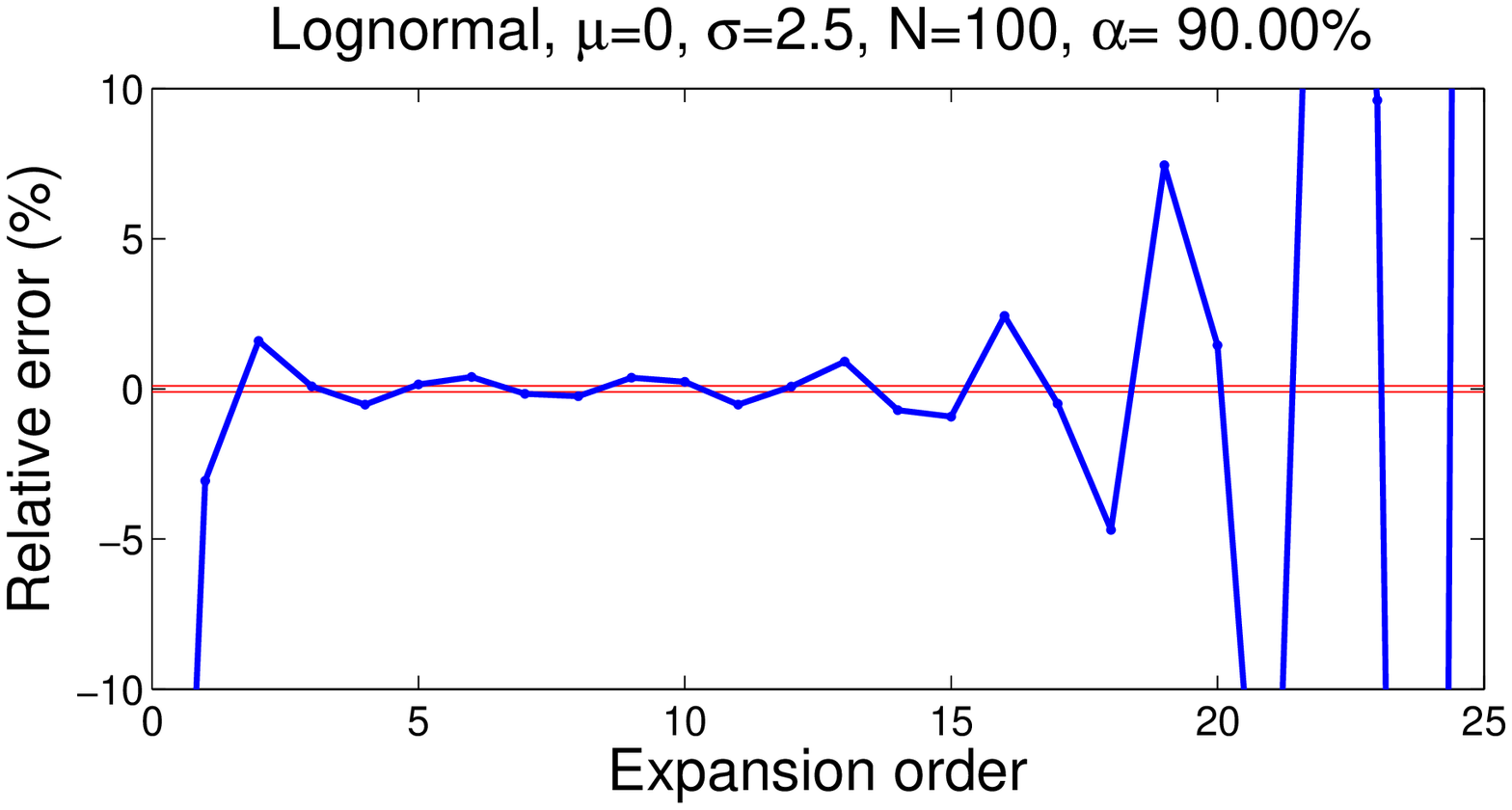} \\
\includegraphics[width=84mm,clip=true]{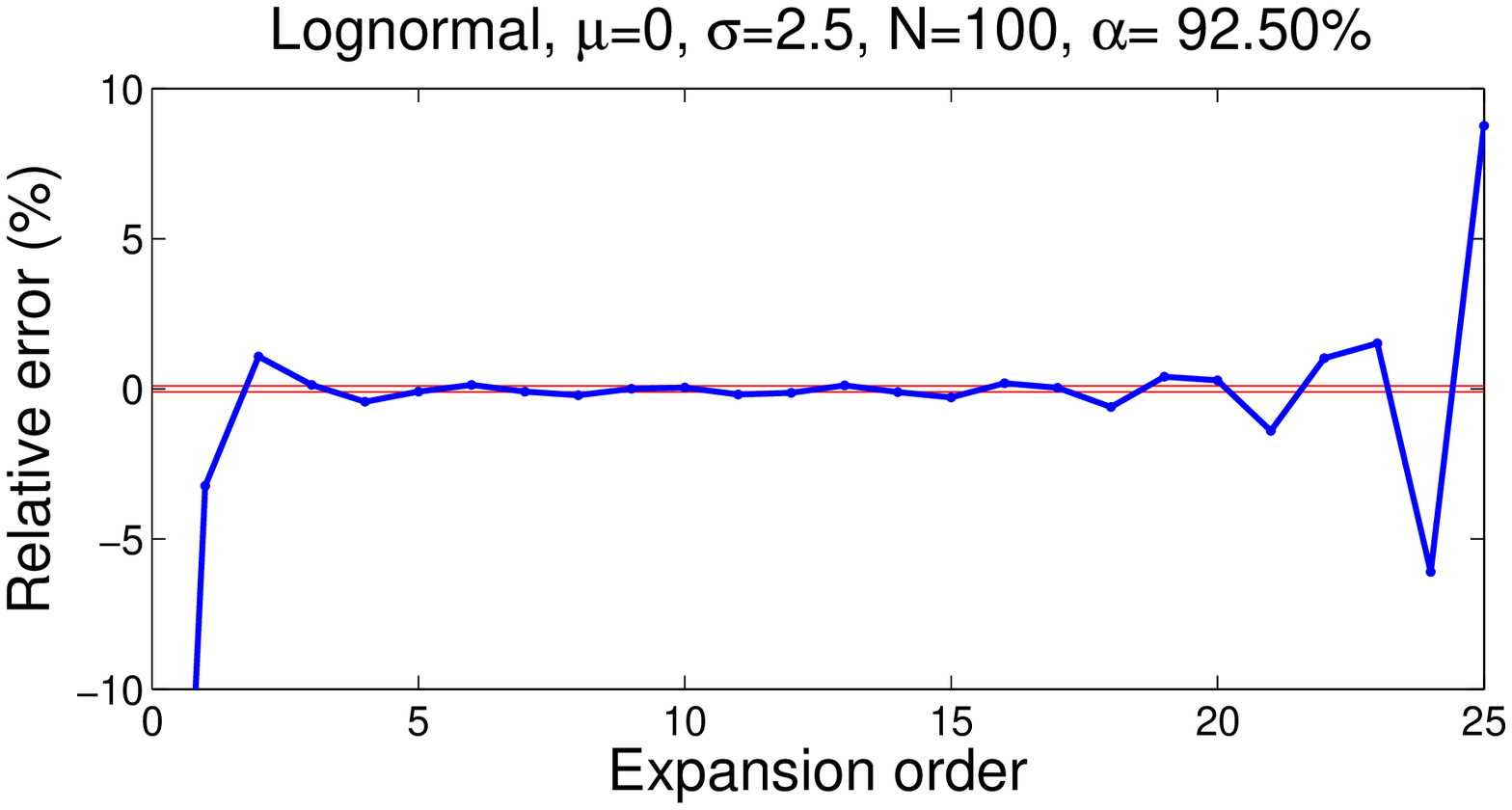} \\
\includegraphics[width=84mm,clip=true]{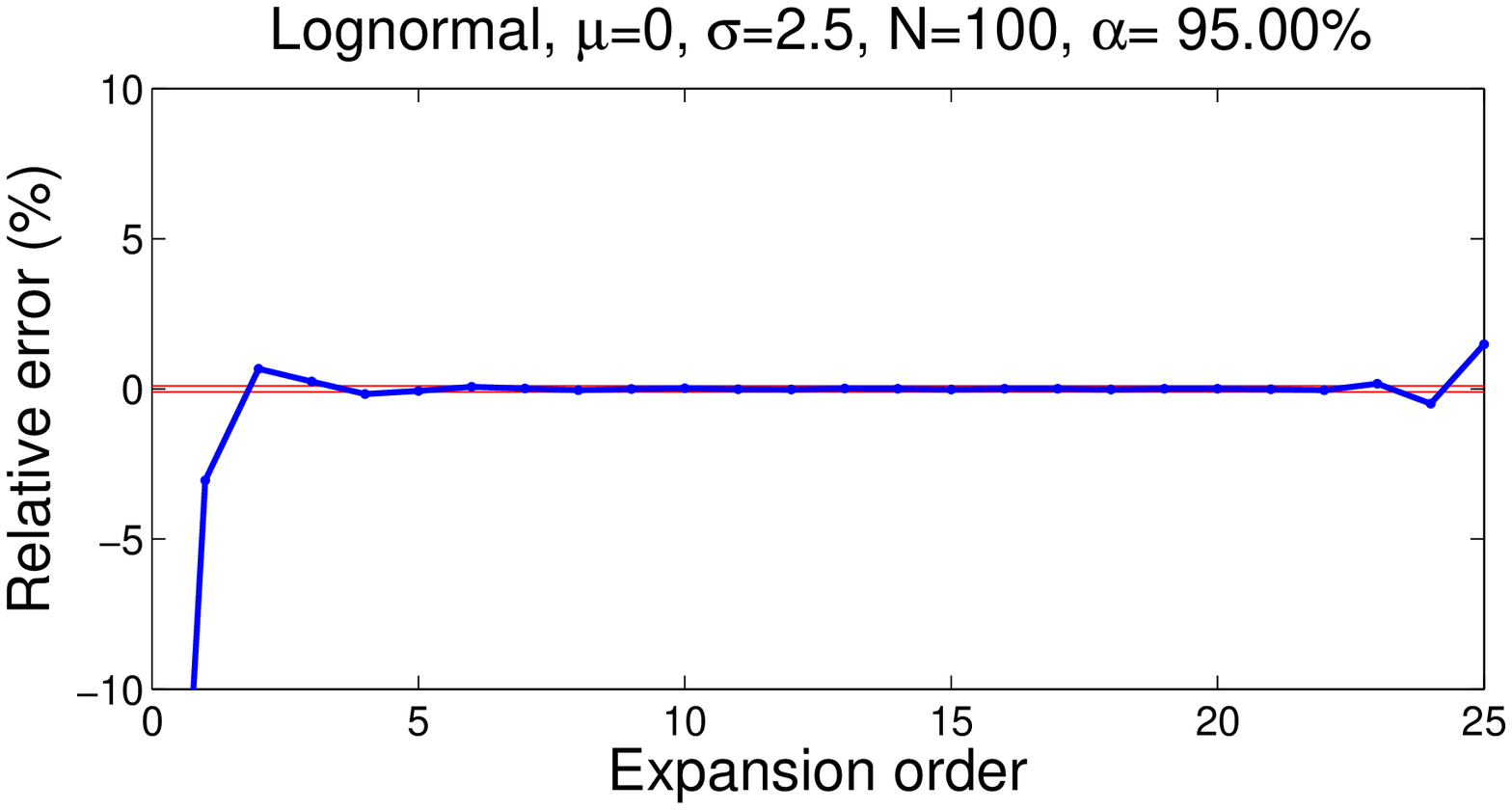} 
\end{tabular}
\end{center}
\caption{{\small Relative error for Lognormal ($\sigma  = 2.5$) with frequency $N=100$, 
as a function of the coefficient order for $ \alpha = 90\% $ (upper plot), 
$ \alpha = 92.5\% $ (middle plot) and $ \alpha = 95\% $ (lower plot). 
The horizontal lines delimit the 95\% confidence interval of the 
Monte Carlo simulation of the exact quantile. }}
\label{fig:convergence1}
\end{figure}

The convergence properties of the perturbative series
are also of great importance. 
Even though a formal analysis of this question
is beyond the scope of this work, we have
carried out an empirical investigation of the accuracy
of the approximation as a function of the 
order at which the perturbative expansion is truncated. 
The results reported are for sums of a fixed number 
of lognormal iidrv's.
Nonetheless, similar patterns are obtained for other 
distributions (e.g. Pareto) in other ranges of
parameters and in sums of iidrv's with random numbers of terms.
In Figure \ref{fig:convergence1}, the relative error
of the quantile estimations for a sum of 
$N=100$ lognormal iidrv's
is plotted as a function of the order of the perturbative expansion, 
for different quantile levels. 
From these results it is apparent that the series converges
only asymptotically for $\alpha \rightarrow 1^{-}$. 
The asymptotic behavior of the series
is analyzed in detail for the particular case
of the Pareto distribution
in section \ref{sssec:expansionParameter}.
For a fixed quantile level, the accuracy of the 
approximation initially improves as more terms are 
included in the expansion, but 
becomes worse beyond a certain order. 
Nonetheless, for a given order, there is a quantile level 
above which the series truncated to this 
order is a more accurate approximation than the 
series truncated to lower orders.
As heavier tails imply stronger dominance of
the maximum in the sum, the heavier the tails of the distribution, 
the more accurate of the approximation becomes. 
Hence, the order beyond which the approximation deteriorates
is larger for distributions with heavier tails.
Finally, the accuracy of the perturbative expansion becomes poorer
for increasing $N$.

In summary, in the cases analyzed, the accuracy 
of the approximation initially improves as more 
terms are included in the perturbative approximation. 
However, beyond a certain order, adding further terms in 
the expansion leads to an increase of the error.
In the experiments carried out in the remainder
of this section, the series is truncated at intermediate orders
($K=3$ or $K=5$), which, for the considered examples, provide very accurate 
approximations. The results of these experiments 
are presented in separate subsections,
each of which corresponds to different 
types of distributions of the individual random variables 
in the sum.

\newpage

\subsection{L\'evy distribution}
In this section we evaluate the accuracy of the different
approximations of high percentiles of the sum of iidrv's that follow 
a L\'evy distribution
\begin{equation}
\begin{aligned}
f(x)  &=  \sqrt{\frac{c}{2 \pi}} \frac{1}{x^{3/2}} e^{-\frac{c}{2x}} \\
F(x)  & = 
\hbox{erfc}\left( \sqrt{ \frac{c}{2x}} \right)   \quad \text{for\ } x > 0,
\end{aligned}
\end{equation}
where $\hbox{erfc}(y)$ is the complementary error function.
The mean of the L\'evy distribution is infinite.
The probability distribution, $F(x)$, is  a function of regular variation $RV_{-a}$
and the  density, $f(x)$, is  $RV_{-(1+a)}$) with $a = 1/2$. 
This a particularly useful case to analyze because the 
L\'evy distribution belongs
to the family of stable distributions \cite{nolan_2012_stable}. 
Therefore, the sum of $N$ L\'evy independent identically 
distributed (iid) random variables 
$ Z_N = \sum_{i=1}^N L_i$,
is also of the L\'evy form 
\begin{equation}
\begin{aligned}
g(z) & =  \sqrt{\frac{c }{2 \pi}} N \frac{1}{z^{3/2}} e^{-\frac{c N^2}{2z}}
 \\
G(z) & =   
\hbox{erfc}\left( \sqrt{ \frac{c}{2z}}  N \right) = 1- \hbox{erf}\left( \sqrt{ \frac{c}{2z}} N \right) 
\end{aligned}
\end{equation}
\begin{figure}[t]
\begin{center}
\begin{tabular}{cc}
\includegraphics[width=84mm,clip=true]{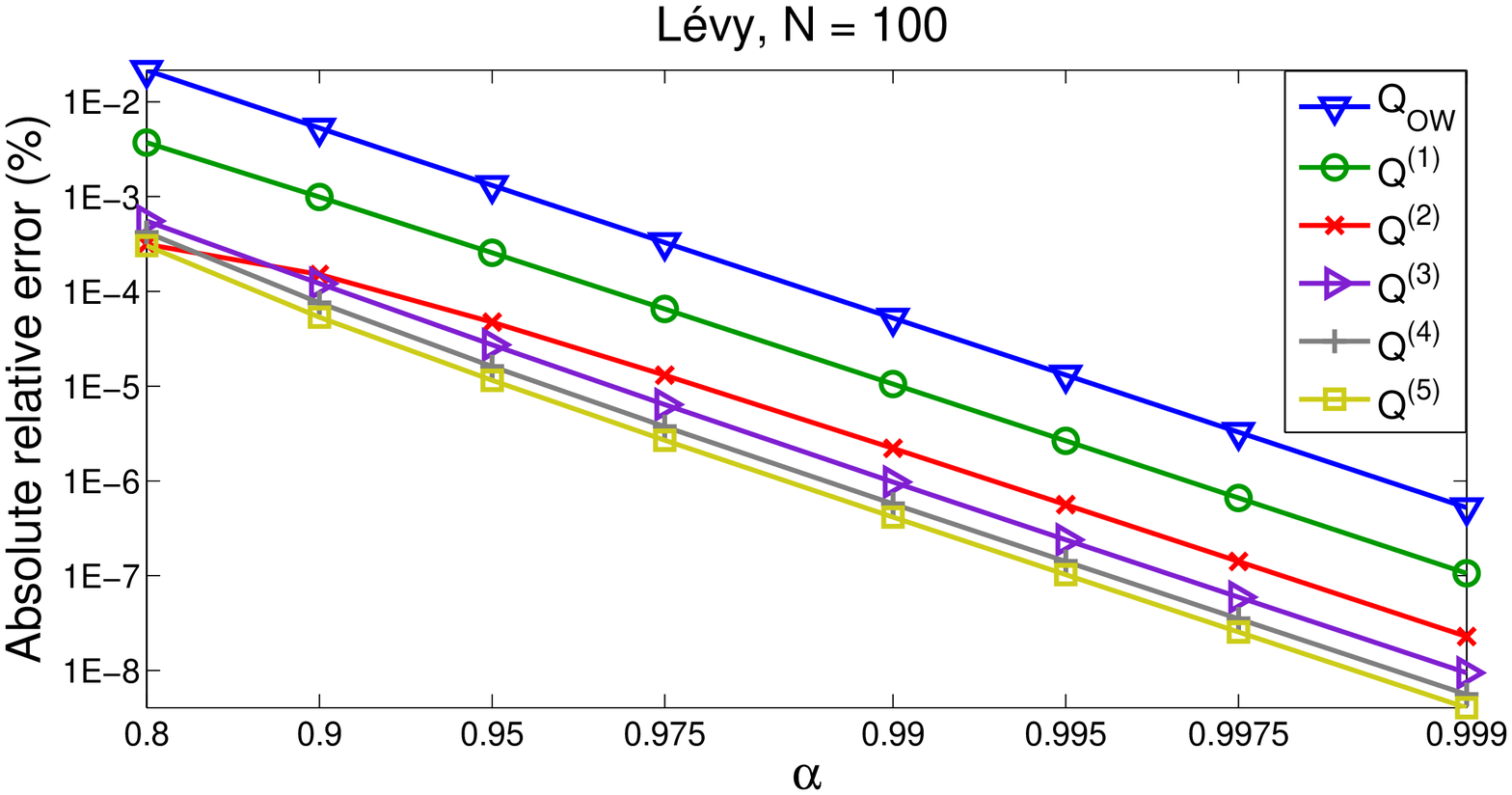} \\
\includegraphics[width=84mm,clip=true]{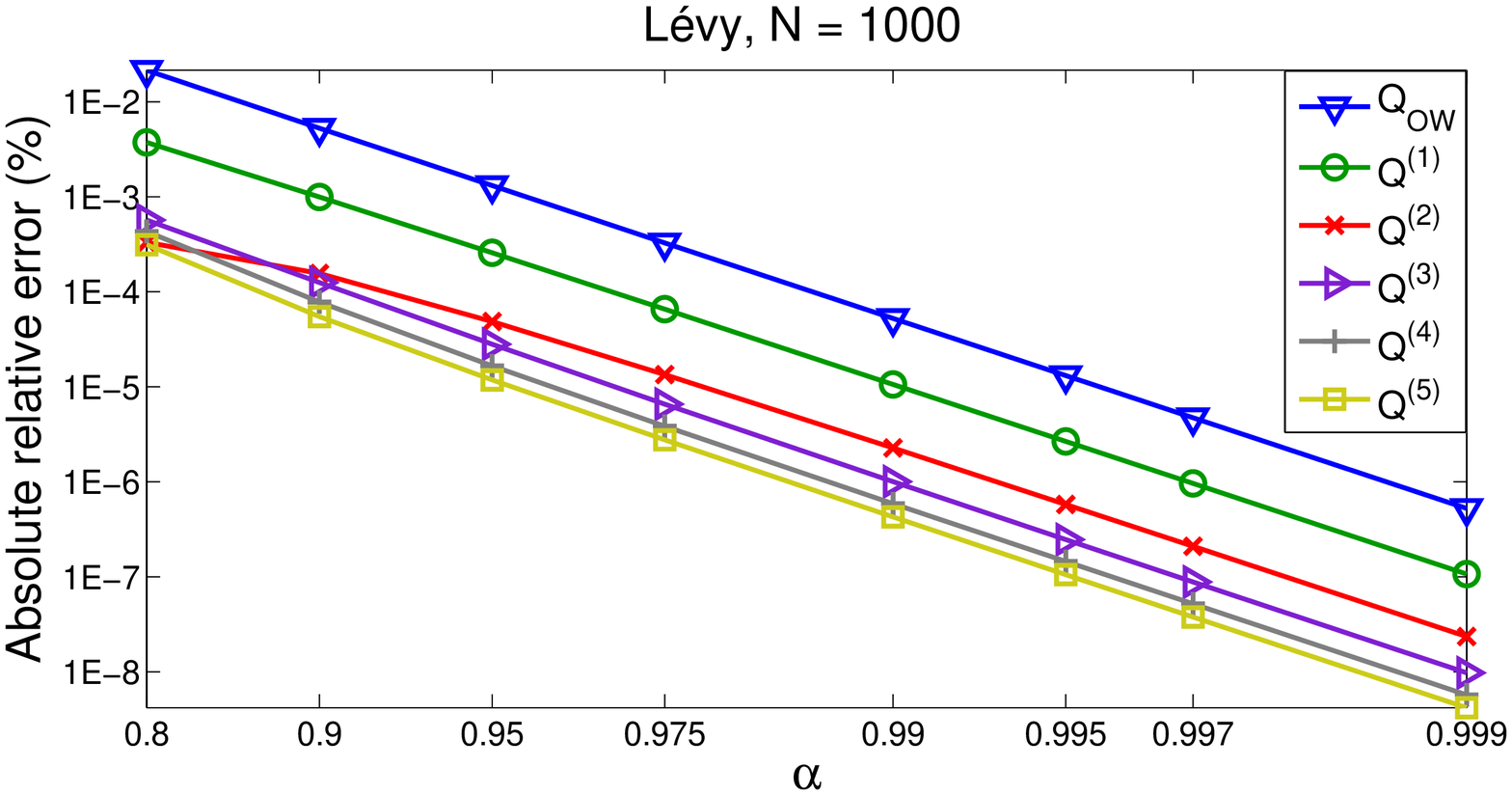}
\end{tabular}
\end{center}
\caption{{\small Absolute value of the  relative error of the 
different approximations  to the $\alpha$-percentile of the 
sum of $N$ independent identically distributed L\'evy random variables 
as a function of $\alpha$  for  $N = 100$ (upper plot) and 
 $N = 1000$ (lower plot).}}
\vspace{-0.5cm}
\label{fig:LEVY_alpha}
\end{figure}

\begin{figure}[t]
\begin{center}
\begin{tabular}{cc}
\includegraphics[width=84mm,clip=true]{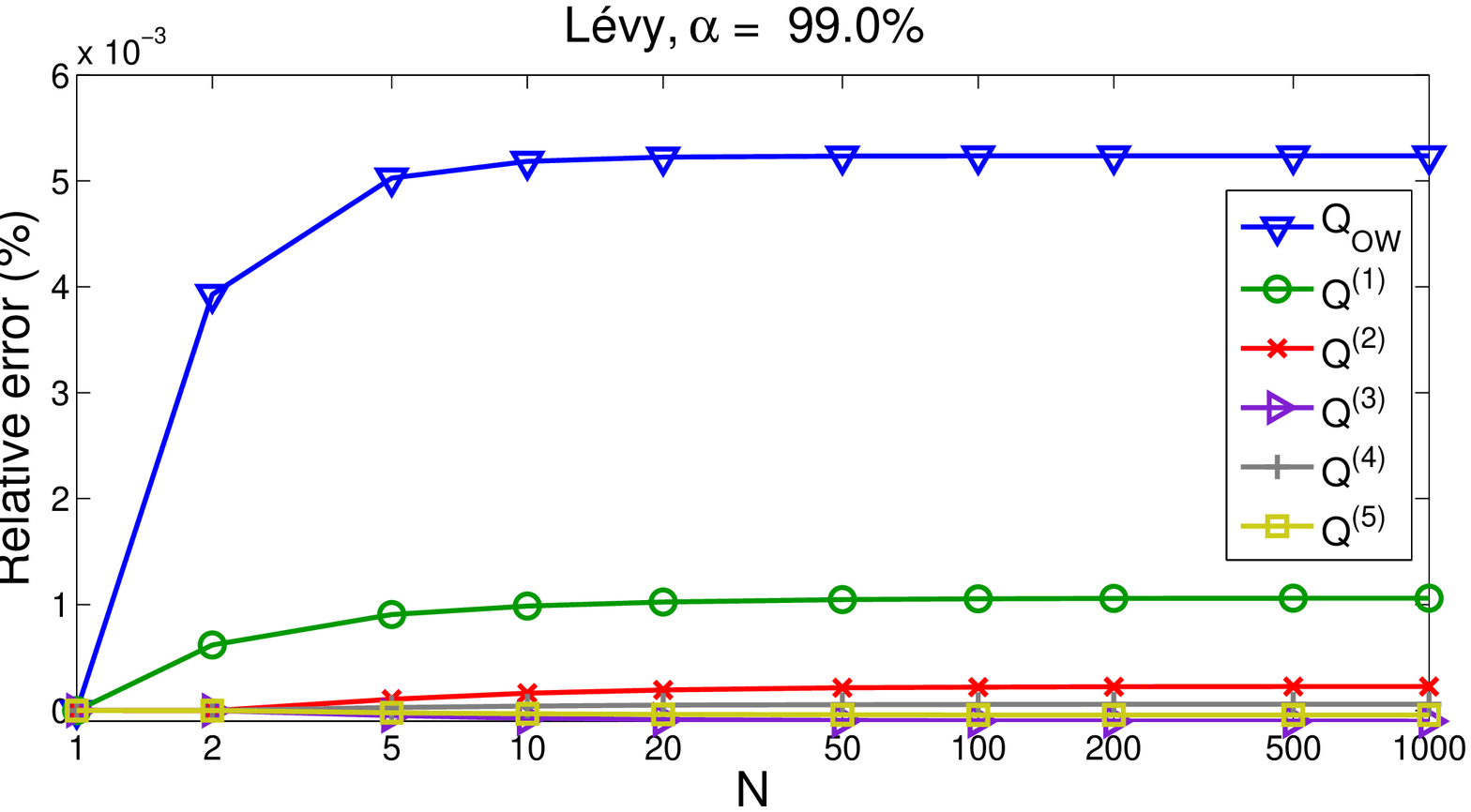} \\
\includegraphics[width=84mm,clip=true]{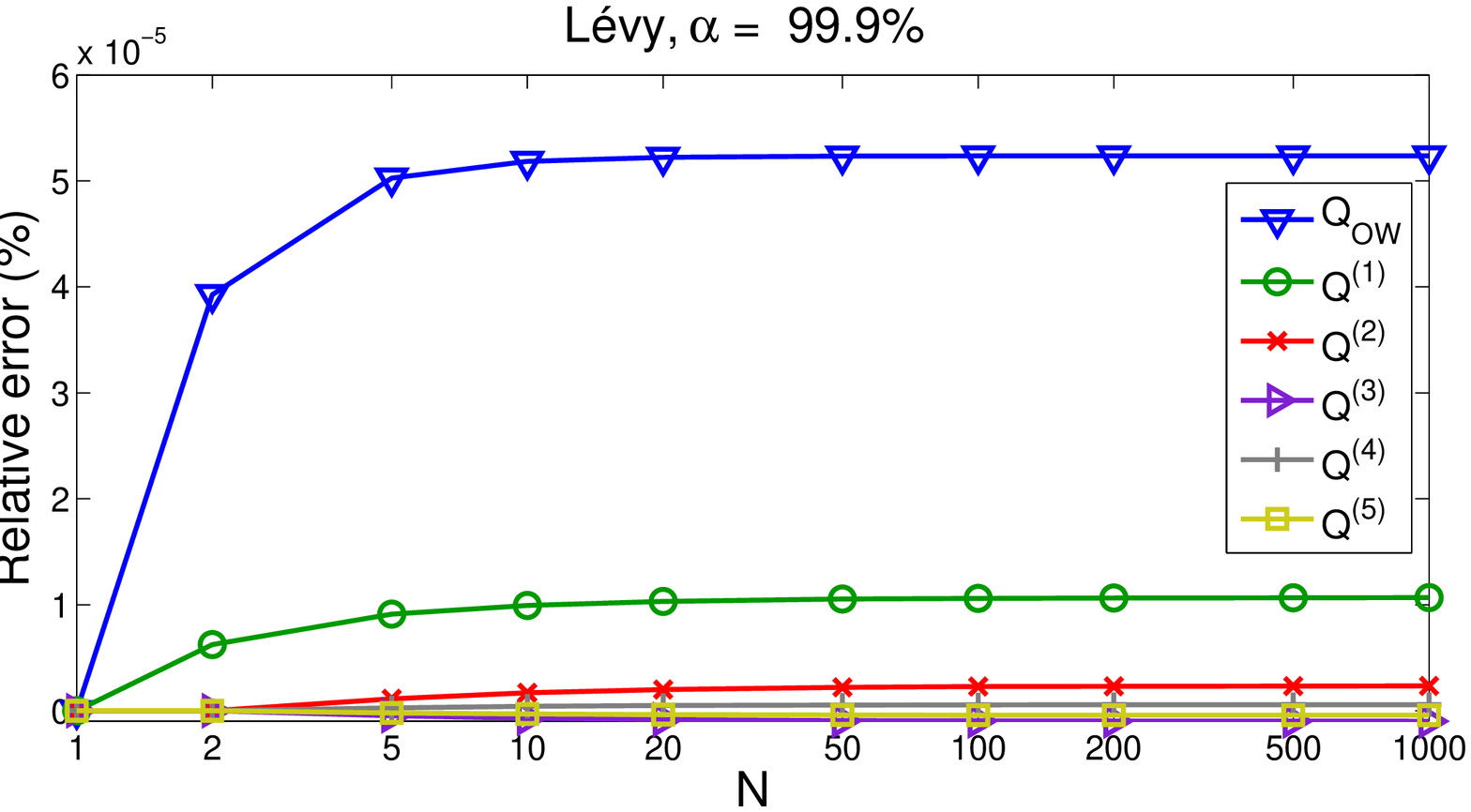}
\end{tabular}
\end{center}
\caption{{\small Relative error of the different approximations 
to  the percentile of 
the sum of iid L\'evy random variables as a function of the number
of terms in the sum for  $ \alpha = 99\% $ (upper plot) and 
 $ \alpha = 99.9\% $ (lower plot).}}
\label{fig:LEVY_N}
\end{figure}
\vspace{-0.5cm}

In the case of deterministic $ N$, the  $\alpha$-percentile is
\begin{equation}
Q = \frac{c}{2} N^2 \left[\hbox{erf}^{-1}\left( 1- \alpha \right) \right]^{-2}.
\end{equation}
For L\'evy random variables,  $c_a = 0$ in (\ref{eq:OW_infiniteMean}) because $a = 1/2$. 
In consequence, the second order asymptotic 
approximation (\ref{eq:OW_infiniteMean})  
coincides with the single-loss approximation
\begin{equation}
\begin{aligned}
\label{eq:OW_Levy} 
Q_{OW} & = Q_{OW}^* = Q_{SL} = F^{-1} \left( 1 - \frac{1 -\alpha}{N} \right)\\
& = \frac{c}{2} \left[\hbox{erf}^{-1}\left( \frac{1 -\alpha}{N} \right) \right]^{-2}. 
\end{aligned}
\end{equation}
The accuracy of this approximation is compared to the 
perturbative series up to order $5$. 
Figure \ref{fig:LEVY_alpha} 
displays in a logarithmic scale in both axes
the absolute value of the relative error of the different
approximations as a function of $\alpha$ 
for $N=100$ and $N=1000$. All approximations become 
more accurate for higher percentiles ($\alpha \rightarrow 1^-$).
In this limit the relative error is proportional to
$ \left(1- \alpha \right)^2 $ for all the approximations considered.
Using the results of Appendix \ref{app:Levy} the
relative error of approximation (\ref{eq:OW_Levy}) is 
\begin{equation}
\frac{Q_{OW} - Q}{Q} \approx \frac{\pi}{6} \frac{N^2-1}{N^2} \left(1- \alpha \right)^2 \quad \alpha \rightarrow 1^-.
\end{equation}
Similarly, for the 
perturbative expansion truncated at different orders
\begin{equation}
\frac{Q^{(k)}-Q}{Q} \approx  \gamma_k \left(1- \alpha\right)^2 \quad \alpha \rightarrow 1^-, \ k  = 1,2,\ldots,
\end{equation}
with
\begin{eqnarray} \label{eq:highPercentiles}
\gamma_1  & =  & \frac{(2\pi -5) N^2 - 6 (\pi-3) N + (4\pi-13)}{12 N^2} \nonumber \\
\gamma_2  & =  & \frac{(N - 1)(N-2)}{6 N^2} \left(\pi-3 \right) \nonumber \\
\gamma_3  & =  &  \frac{(N-1)(N-2)}{6 N^2} \left(\pi-\frac{16}{5}\right). 
\end{eqnarray}

Up to the orders analyzed the perturbative 
series provides more accurate estimates than
(\ref{eq:OW_Levy}), improving 
with the number of terms included in this series. Nonetheless,
the relative improvements become smaller for higher order terms.
The dependence of the relative error with
$N$, the number of terms in the
sum, for $ \alpha = 99\% $ (upper plot) and 
 $ \alpha = 99.9\% $ (lower plot) 
is shown in Figure  \ref{fig:LEVY_N}.
The relative error increases with $N$.
Nonetheless, the deterioration is fairly slow.
The error eventually approaches a constant, in agreement 
with the large $N$ behavior of (\ref{eq:highPercentiles}). 
Also in these cases the perturbative series is more accurate 
that $Q_{OW}$.

\subsection{Lognormal distribution}

In this section we analyze the sum of iidrv's that 
follow a lognormal distribution
\vspace{0.25cm}
\begin{equation}
\begin{aligned}
f(x)  &=  \frac{1}{x \sigma \sqrt{2 \pi}} 
\exp \left( - \frac{(\log x)^2}{2 \sigma^2} \right)\\
F(x)  &= \frac{1}{2} + \frac{1}{2} 
\hbox{erfc}\left( \frac{\log x}{\sigma \sqrt{2}} \right),   
\quad \text{for\ } x > 0.
\end{aligned}
\end{equation}
The lognormal  is also subexponential. However, in contrast
to the L\'evy distribution, all its moments are finite. 
The perturbative series, which is of the same form as
in the previous case, also provides very accurate approximations
of high percentiles of the sum.

\begin{figure}[t]
\begin{center}
\begin{tabular}{cc}
\includegraphics[width=84mm,clip=true]{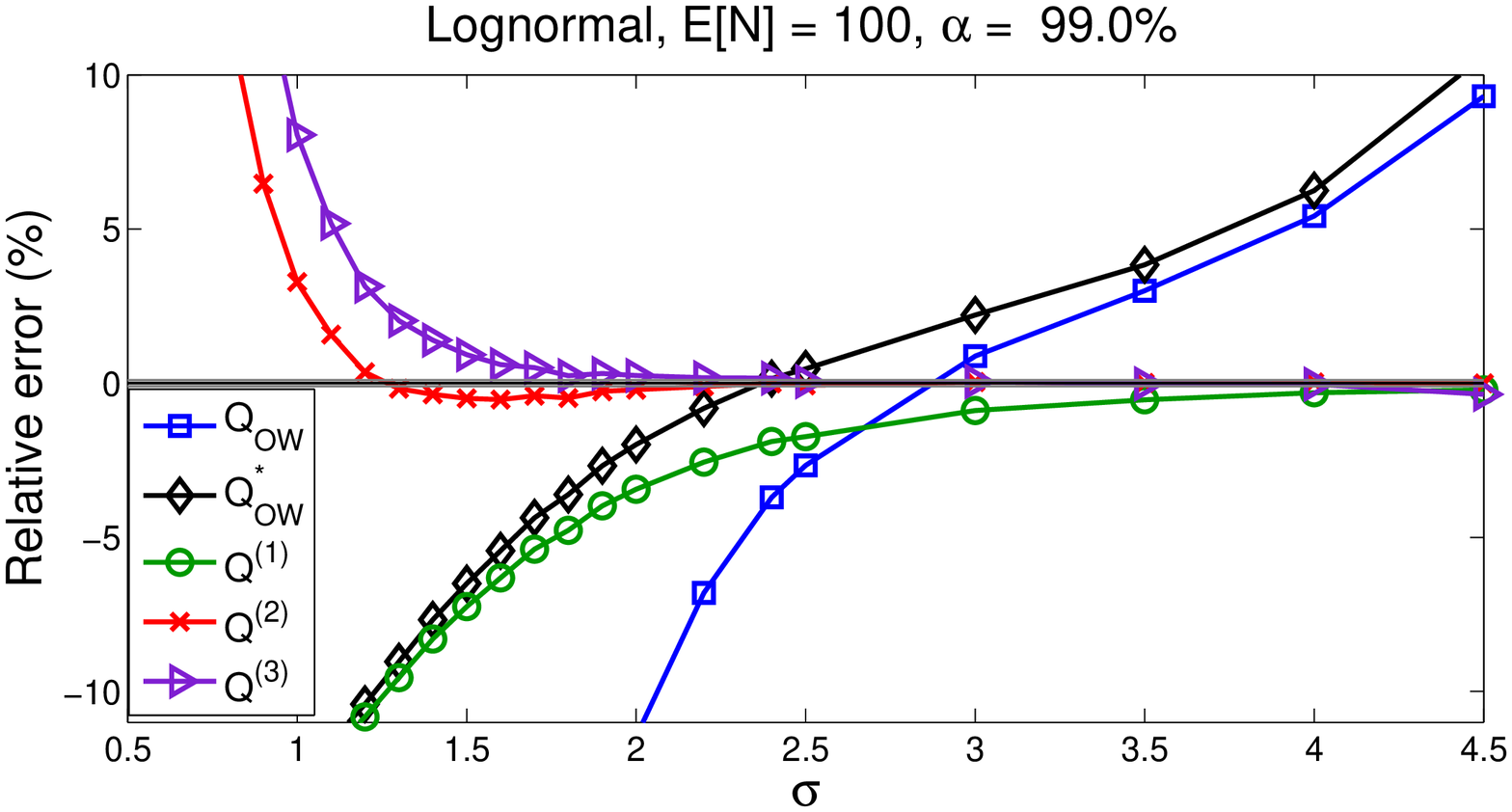} \\
\includegraphics[width=84mm,clip=true]{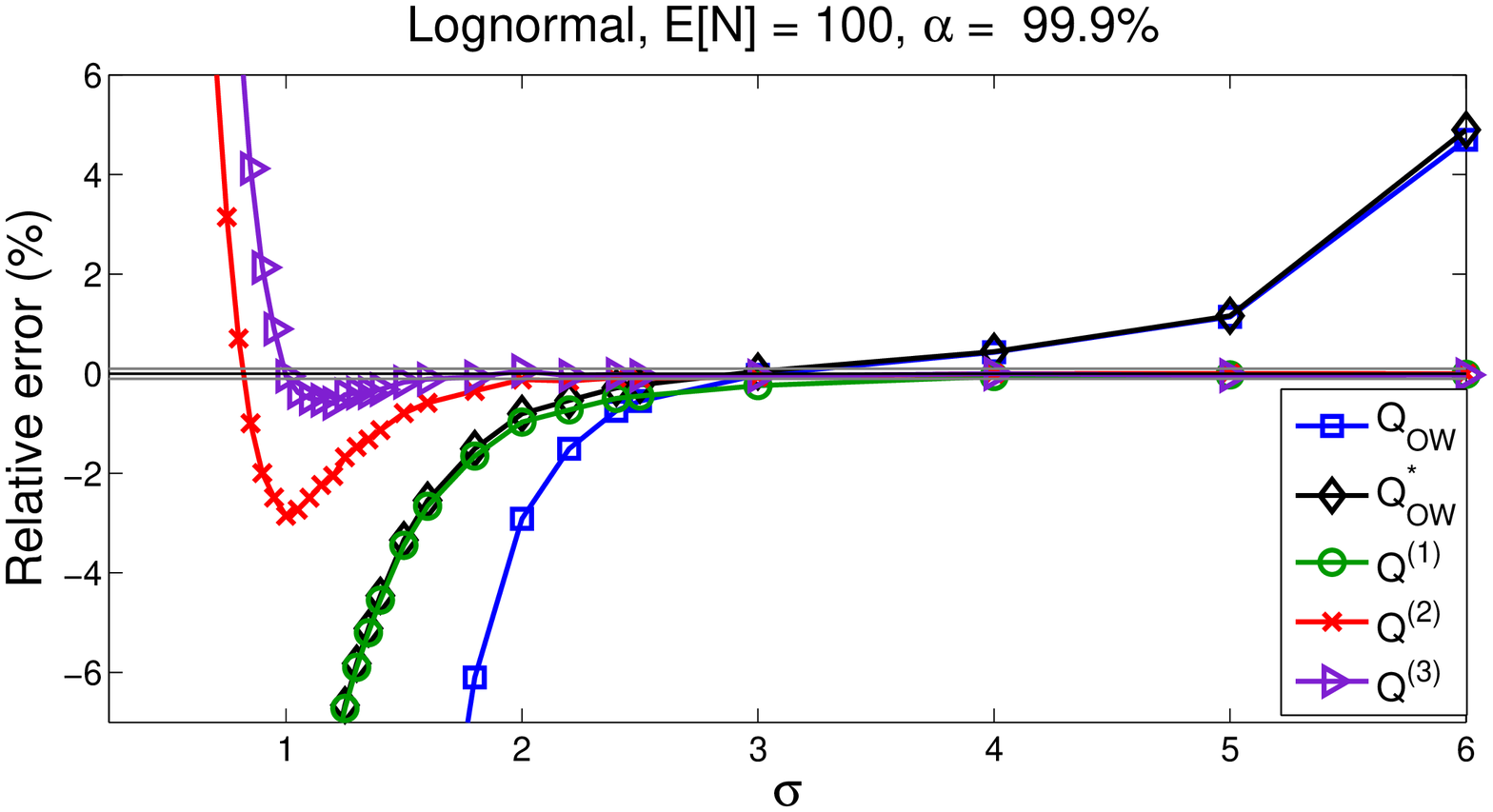}
\end{tabular}
\end{center}
\caption{{\small Relative error for Lognormal /Poisson ($\lambda  =100$) 
as a function of $\sigma$ for  $ \alpha = 99\% $ (upper plot) and 
 $ \alpha = 99.9\% $ (lower plot).}}
\label{fig:LN_sigma}
\end{figure}

Figure \ref{fig:LN_sigma} displays the relative error of the
different approximations as a function of $\sigma$.
Larger values of $\sigma$ correspond to heavier tails. 
In the simulations the number of terms in the sum (frequency) 
is random and follows a Poisson distribution whose mean is $ \lambda=100 $.  
In all cases, the relative error becomes smaller as $ \sigma $  increases. 
This is  consistent with the fact that this parameter determines 
the heaviness of the tail. For larger values of $\sigma$
(heavier tails) the relative importance of the maximum in
the sum increases and the approximations, which 
are based on the dominance of the maximum in the sum, become more accurate. 

\begin{figure}[!t]
\begin{center}
\begin{tabular}{cc}
\includegraphics[width=84mm,clip=true]{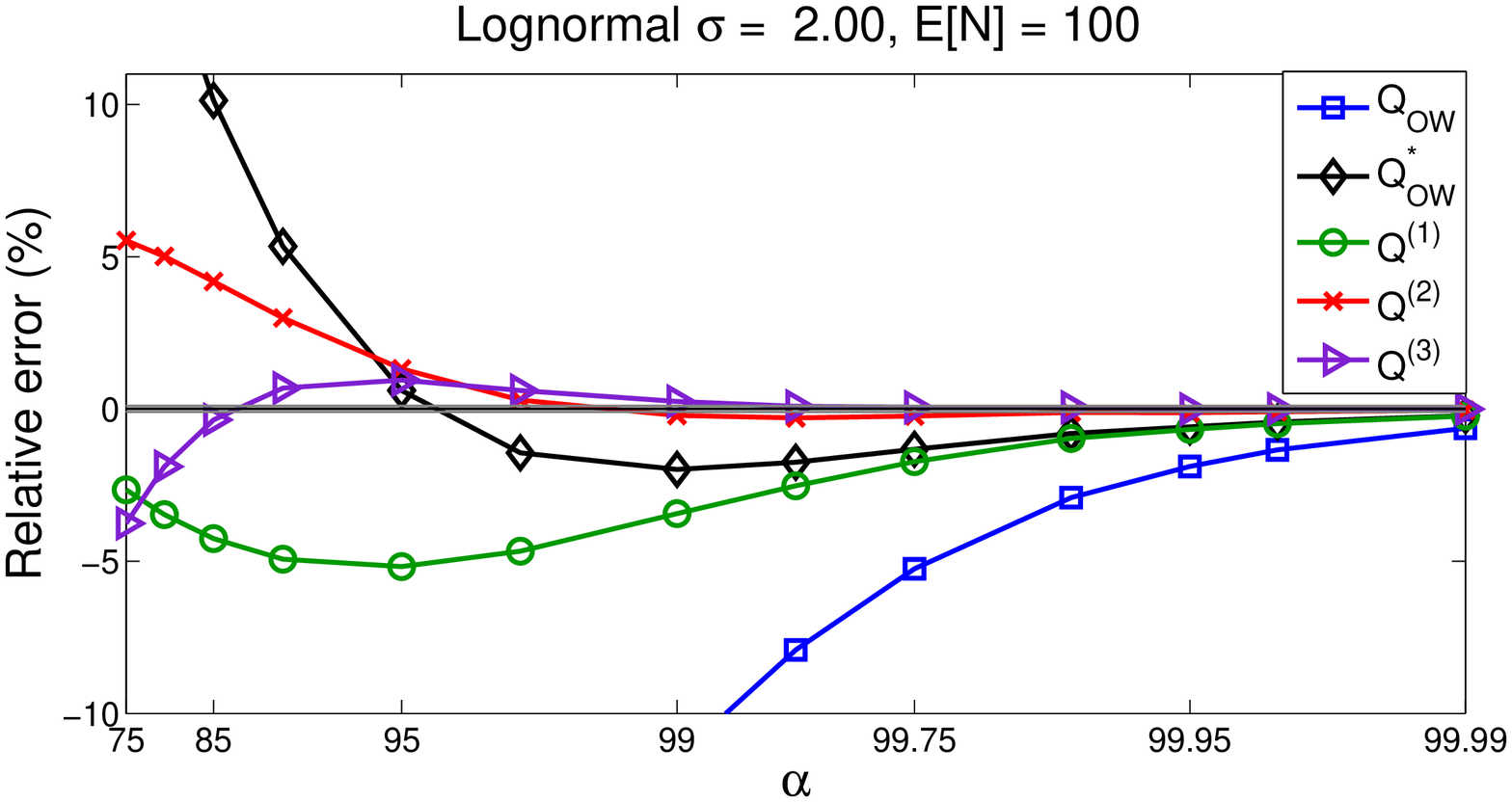} \\
\includegraphics[width=84mm,clip=true]{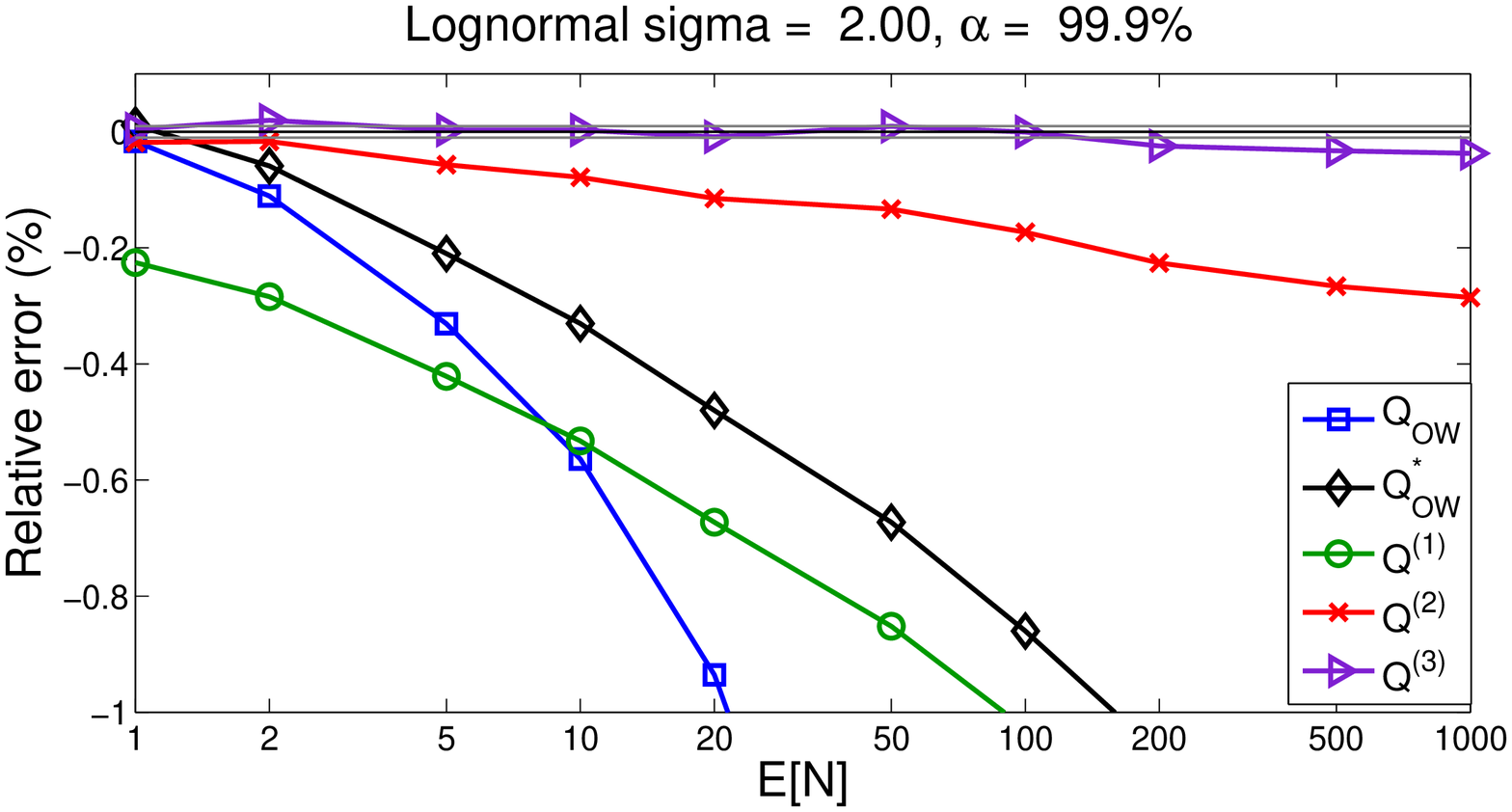}
\end{tabular}
\end{center}
\caption{{\small Relative error for Lognormal($\sigma = 2$)/Poisson  
as a function of $\alpha$ for $\mathtt{E}[N] = 100$ (upper plot) and 
as a function of $\mathtt{E}[N]$ for $ \alpha = 99.9\% $ (lower plot)}}
\label{fig:LN_alpha_N}
\end{figure}

The second order asymptotic approximations $ Q_{OW}^* $ and $ Q_{OW} $ 
diverge as $\sigma$ becomes larger. This is not unexpected because
the mean of the distribution increases as $e^{\sigma^2/2}$, 
while the percentile of the maximum (which dominates the sum) 
increases only as $e^{\sigma}$. The perturbative expansion 
introduced in this work, which 
involves only \textit{censored} moments, avoids this problem and 
behaves properly.  
Figure \ref{fig:LN_alpha_N} displays the dependence of the error of the 
different approximations as a function of the percentile level (upper plot) and 
of the average frequency (lower plot).
As expected, all approximations perform better at higher percentiles 
and lower frequencies; that is, as the weight of the maximum in the sum 
becomes larger. Even for the relatively 
high average frequency $\lambda=1000$, the accuracy 
of the perturbative approximation  $Q^{(3)}$ is remarkable.

\subsection{Pareto distribution}

In this section we analyze the sum of iidrv's that follow a Pareto distribution
\begin{eqnarray}
f(x) =  \frac{a}{x^{1+a}},  \quad F(x)  =  1 - \frac{1}{x^a},  \quad  x > 1, 
\end{eqnarray}
with $a >0$. Since the second order asymptotic approximations have a 
different form depending on whether the mean is defined or not, 
we consider two separate regimes: $ a > 1 $, where 
the mean of the Pareto distribution is finite, and $ a \leq 1 $,
 where the mean diverges. 
It is worth noting that the perturbative expansion introduced
in this work has the same expression in both regimes and 
is in fact continuous at $a=1$.

\begin{figure}[t]
\begin{center}
\begin{tabular}{cc}
\includegraphics[width=84mm,clip=true]{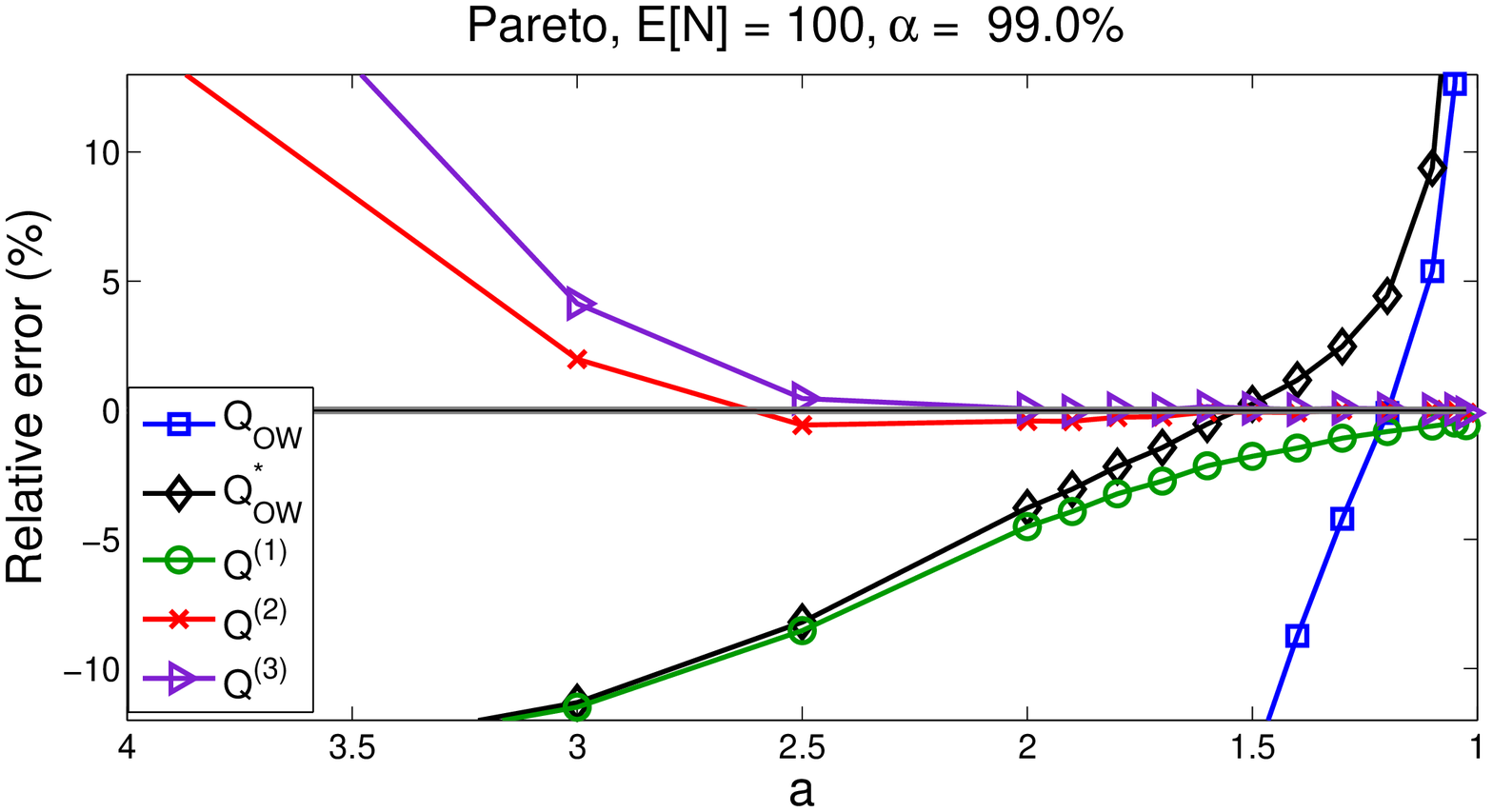} \\
\includegraphics[width=84mm,clip=true]{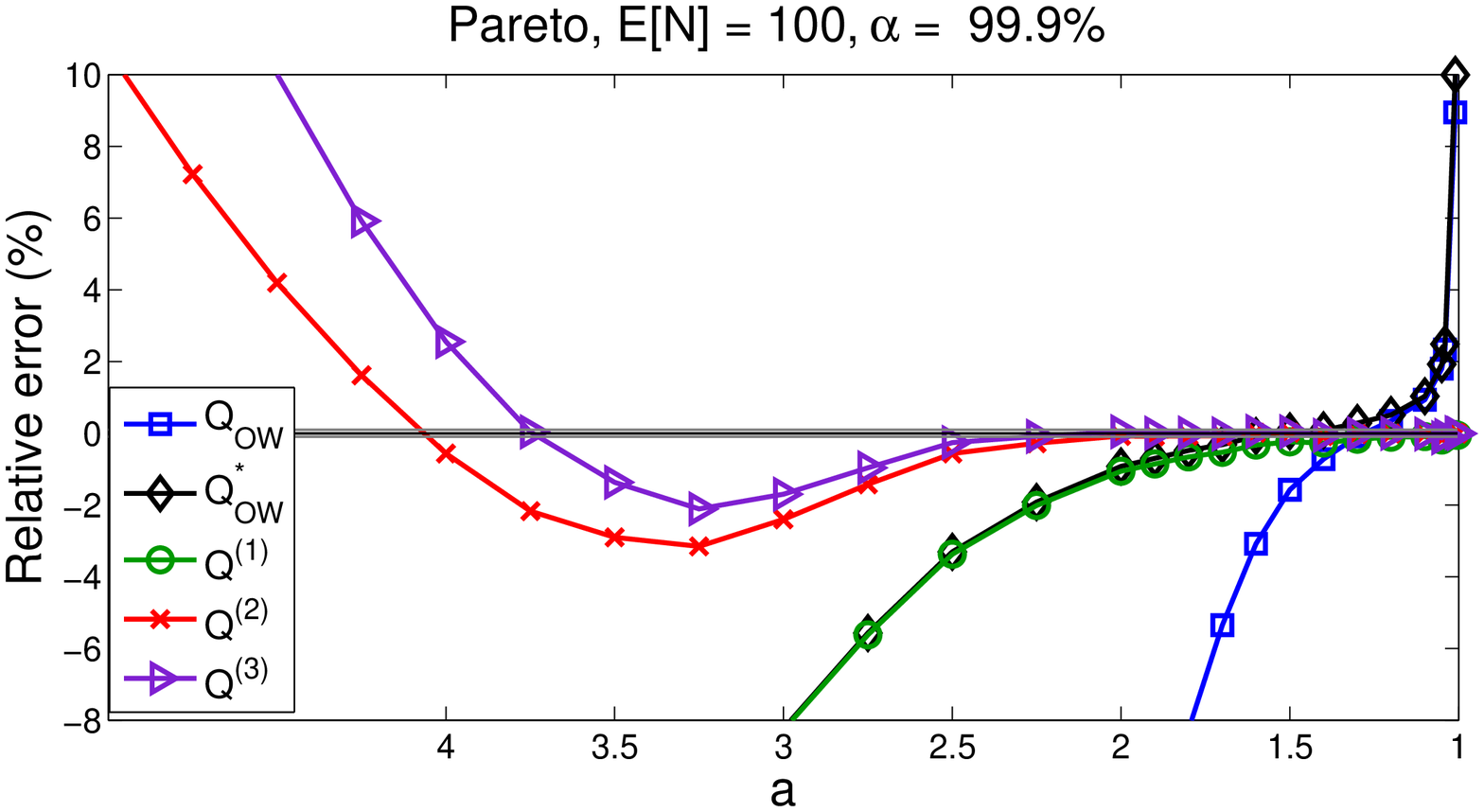} 
\end{tabular}
\end{center}
\caption{{\small Relative error for Pareto / Poisson ($\lambda  = 100$) 
as a function of $a$ for  $ \alpha = 99\% $ (upper plot) and 
 $ \alpha = 99.9\% $ (lower plot).  The values of $a$ are ordered so that the
 heaviness of the tails increases from left to right in the plots.}}
\label{fig:pareto_a_finite}
\end{figure}

\begin{figure}[!h]
\begin{center}
\begin{tabular}{cc}
\includegraphics[width=84mm,clip=true]{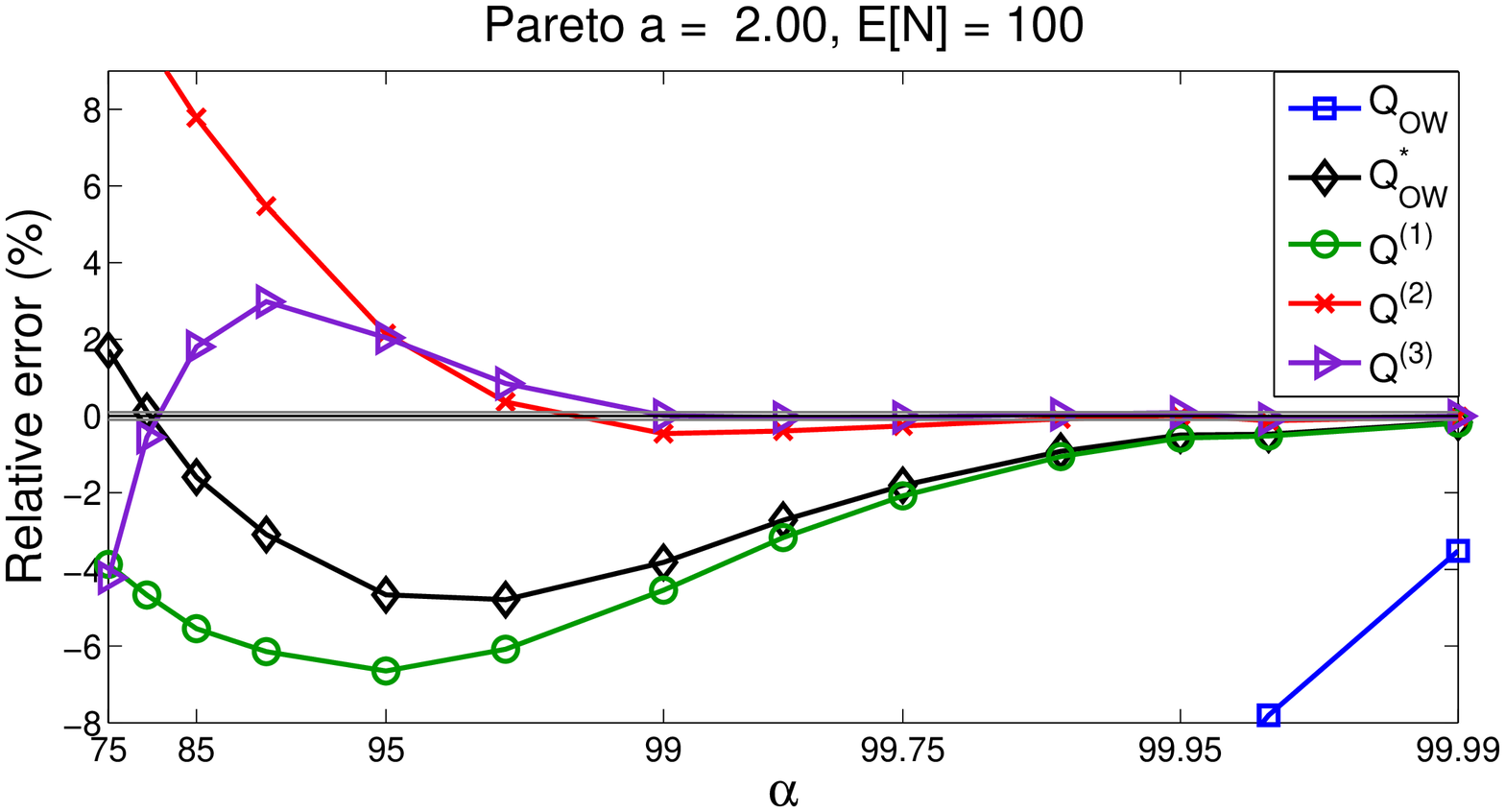} \\  
\includegraphics[width=84mm,clip=true]{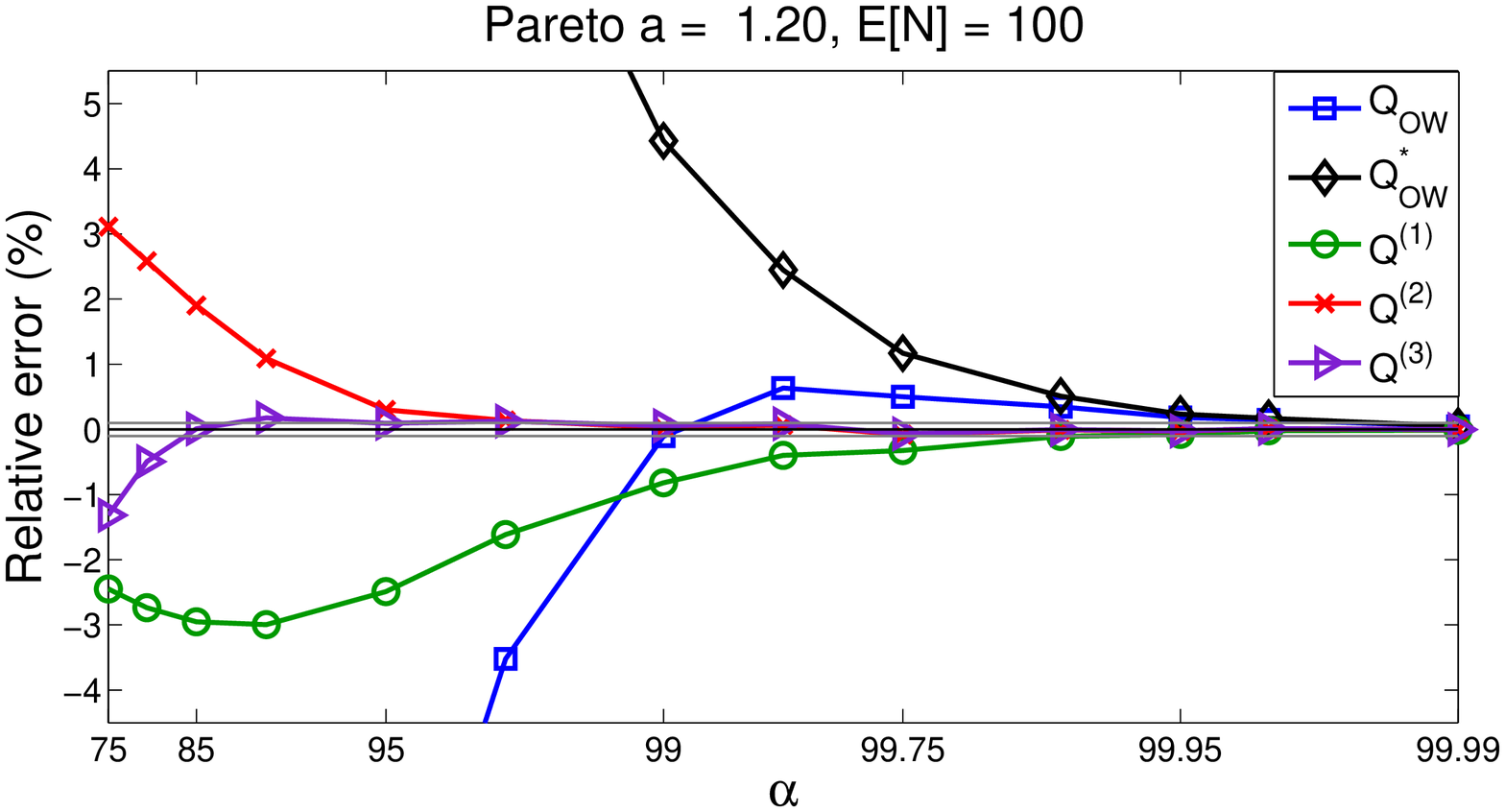}
\end{tabular}
\end{center}
\caption{{\small Relative error for Pareto / Poisson ( $\lambda = 100$ )   
as a function of $\alpha$ for different values of $a$: 
$a = 2.00$ (upper plot) and $ a = 1.20$ (lower plot).}}
\label{fig:pareto_alpha_finite}
\end{figure}

\begin{figure}[t]
\begin{center}
\begin{tabular}{cc}
\includegraphics[width=84mm,clip=true]{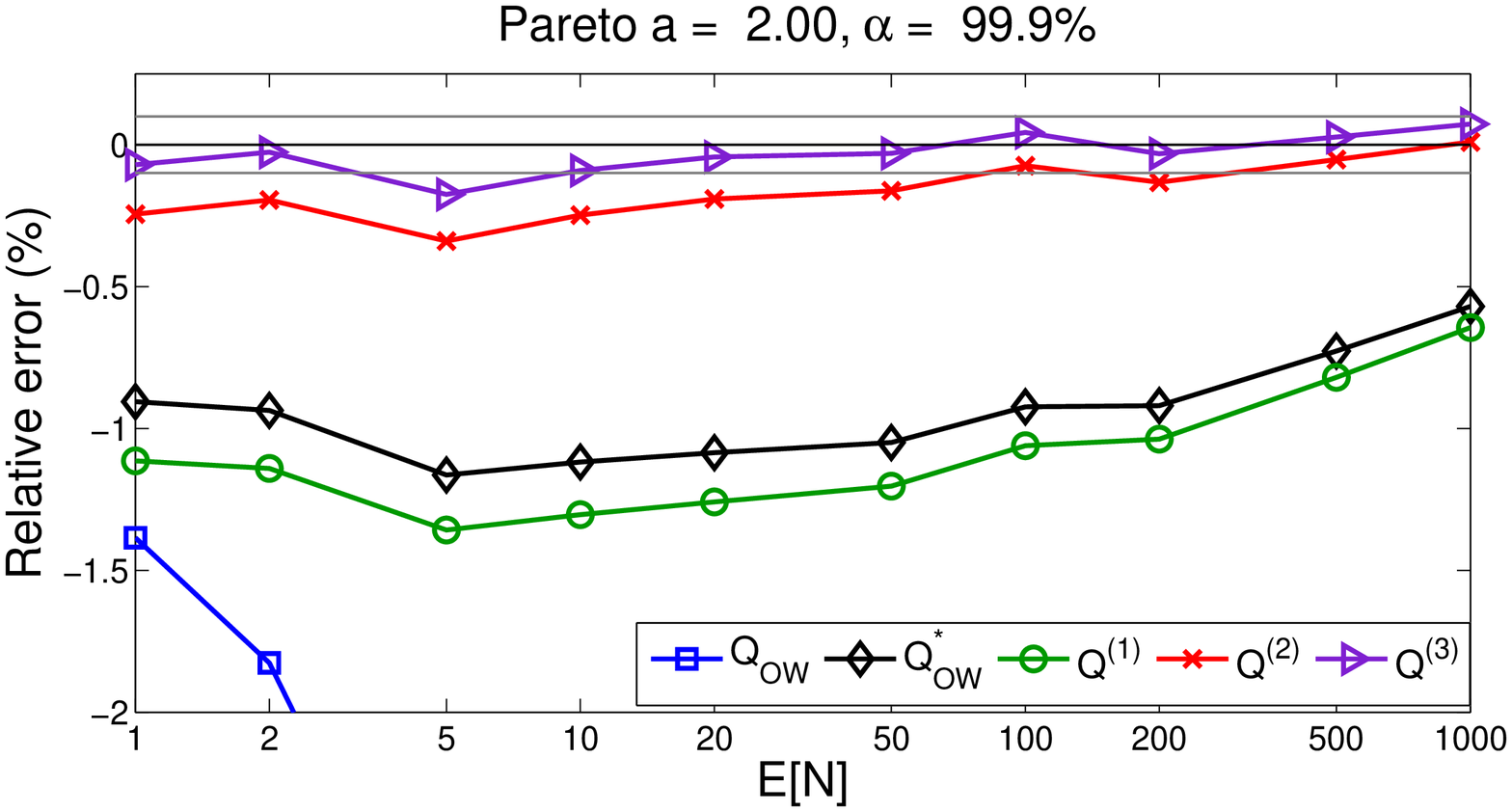} \\
\includegraphics[width=84mm,clip=true]{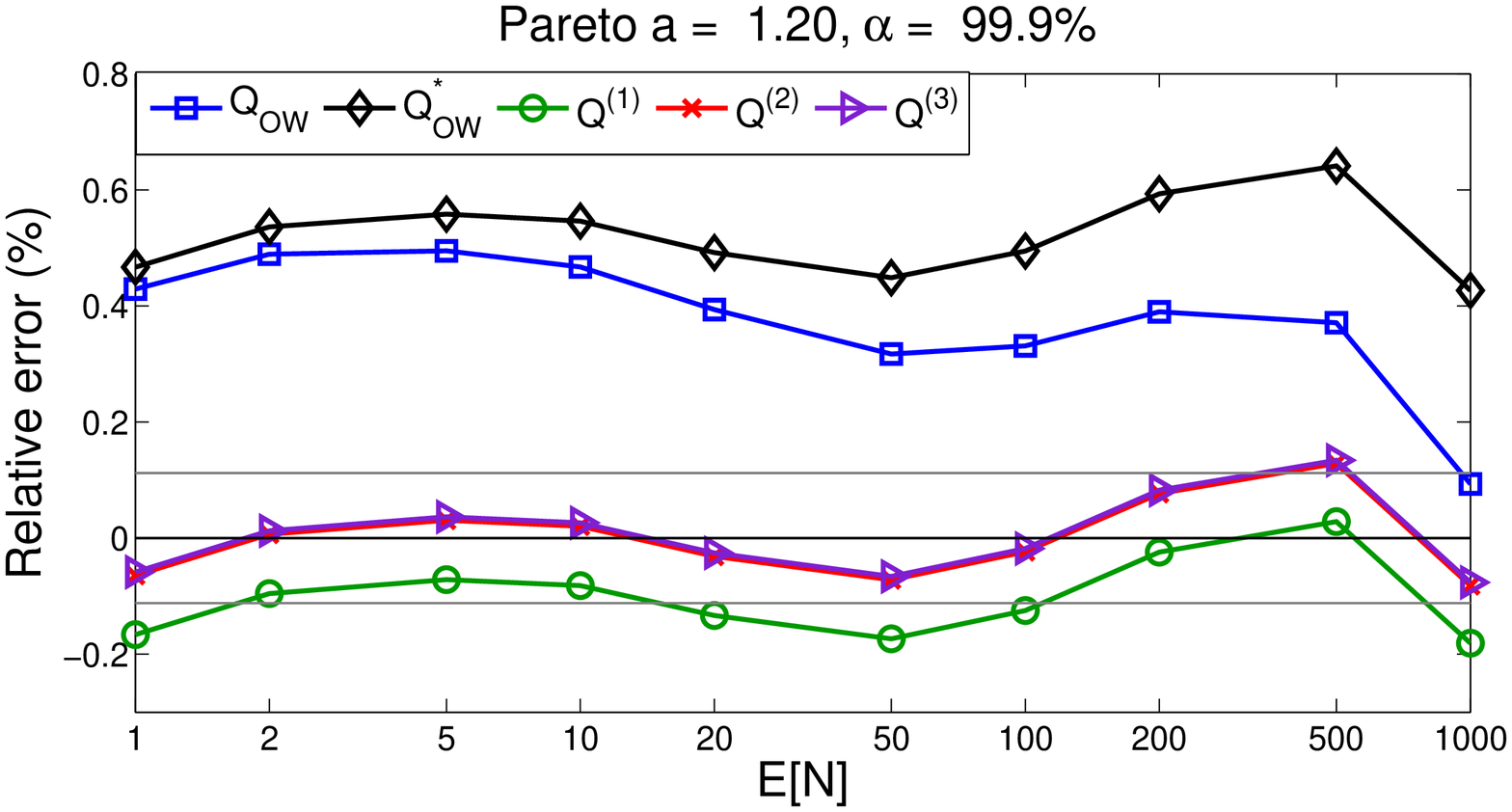}
\end{tabular}
\end{center}
\caption{{\small Relative error for Pareto/Poisson  
as a function of $\EX[N]$ for $\alpha = 99.9\% $
and different values of $a$: $a = 2.00$ (upper plot) 
and $ a = 1.20$ (lower plot).}}
\label{fig:pareto_lambda_finite}
\end{figure}

\subsubsection{Pareto distribution with finite mean $(a>1)$:}

We now compare the accuracy of the different
approximations for sums of random variables
that follow a Pareto distribution with finite mean
using Monte Carlo simulations.
Figure \ref{fig:pareto_a_finite} displays the 
relative error as
a function of the Pareto index $a$. 
In the limit $a\rightarrow 1^+$
the second order asymptotic approximations $Q_{OW}$ and $Q_{OW}^*$ 
diverge. The origin of this divergence is the
increase of correction term in  
(\ref{eq:OW_finiteMean},\ref{eq:OW_finiteMean_perturbative}),
which involves the unconditional mean of the distribution. 
This mean which grows without bound as $a$ approaches $1$ 
from above.
By contrast, the perturbative expansion, which is expressed
in terms of censored moments, behaves well 
and actually becomes more accurate in this limit.
Figure \ref{fig:pareto_alpha_finite} presents the dependence of
the relative error as a function of
$\alpha$. The dependence on the average 
frequency  $ \lambda = \EX[N] $ is shown in Figure \ref{fig:pareto_lambda_finite}.
In all cases the conclusions reached through the analysis of these results
are similar to the lognormal case. 

\subsubsection{Pareto distribution with infinite mean $ ( 0  < a \le 1) $:}
\begin{figure}[!ht]
\begin{center}
\begin{tabular}{cc}
\includegraphics[width=84mm,clip=true]{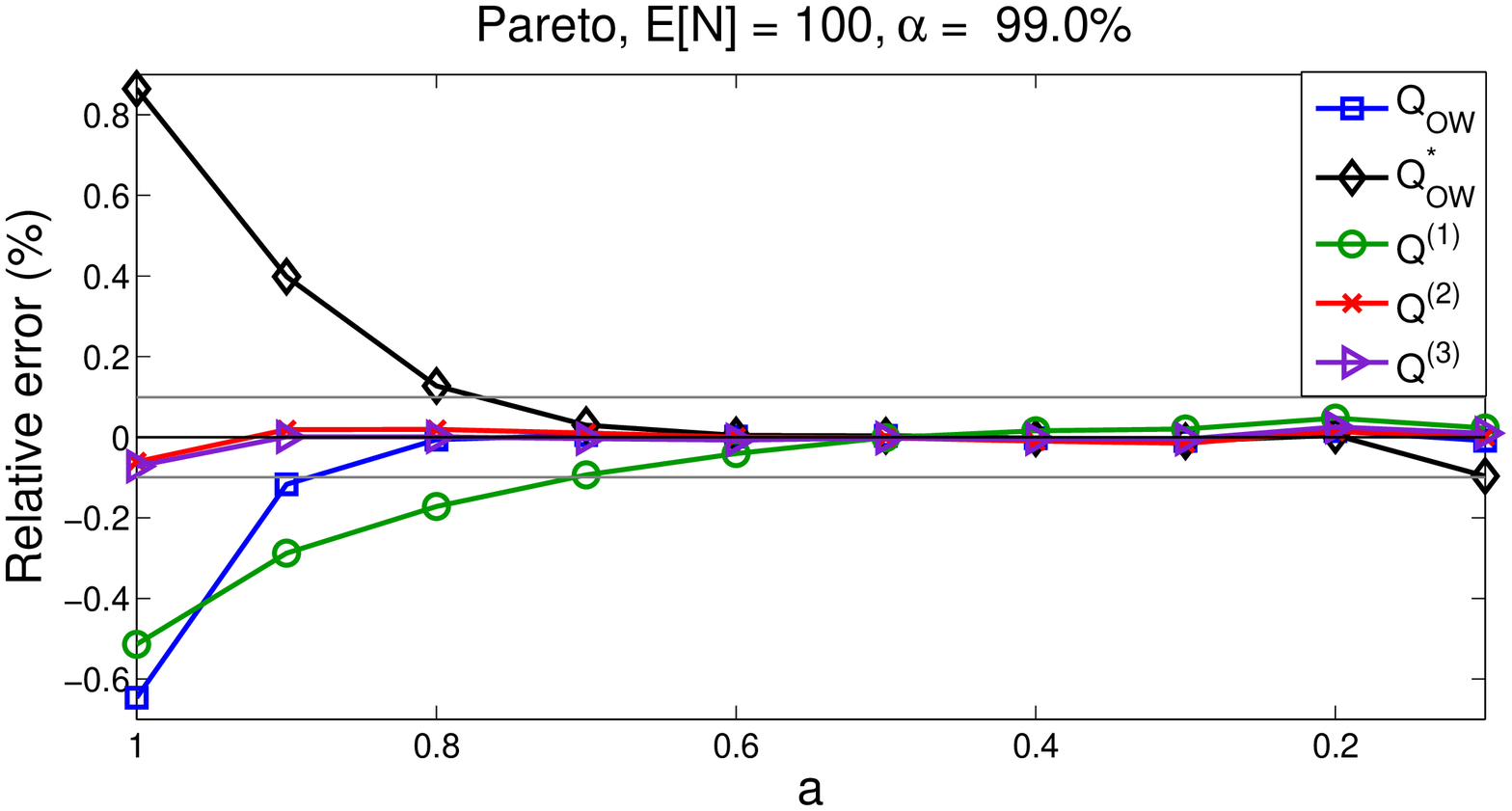} \\
\includegraphics[width=84mm,clip=true]{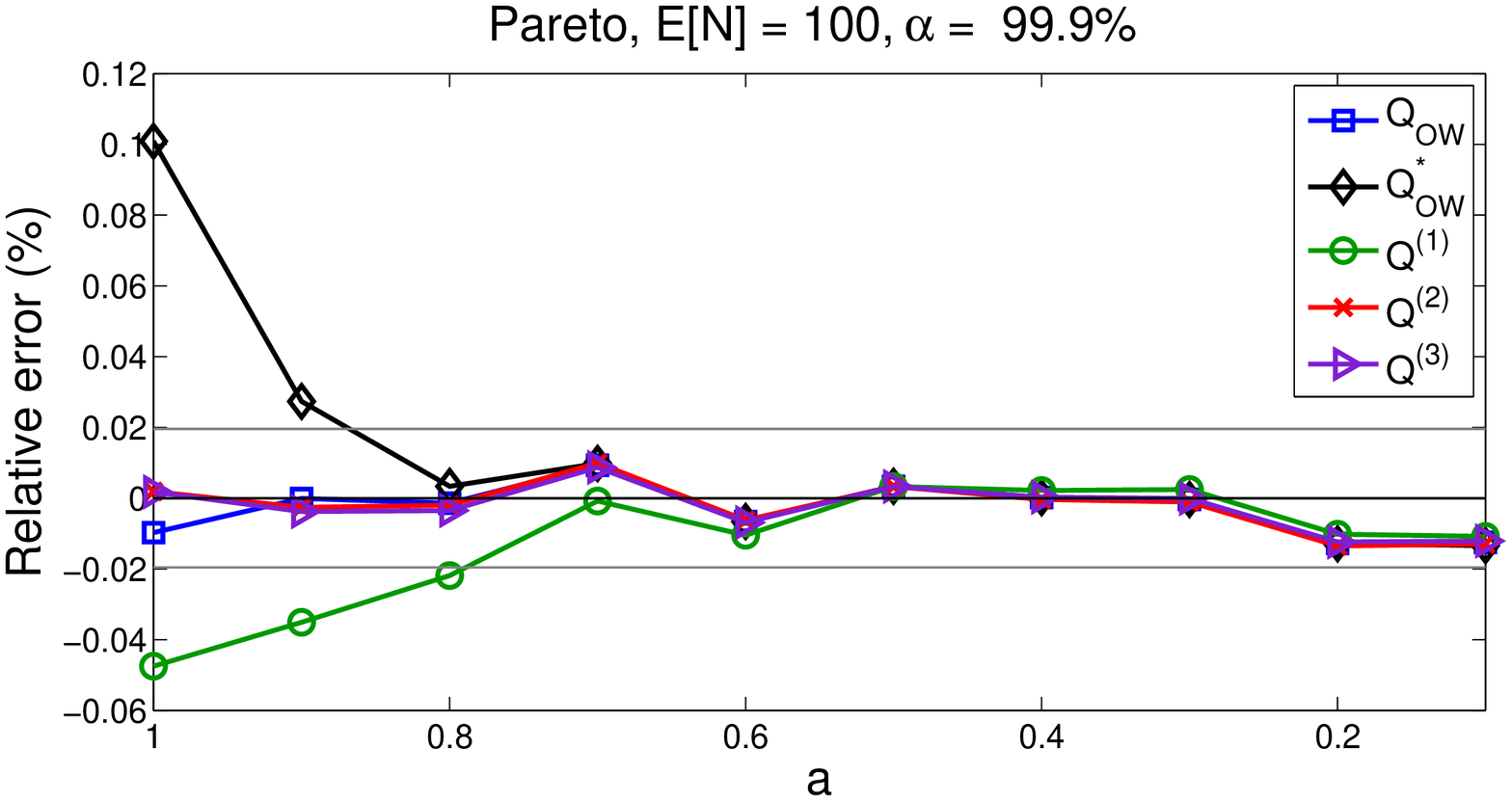} 
\end{tabular}
\end{center}
\caption{{\small Relative error for Pareto/Poisson ($\lambda  = 100$) 
as a function of $a$ for  $ \alpha = 99\% $ (upper plot) and 
 $ \alpha = 99.9\% $ (lower plot). The values of $a$ are ordered so that the
 heaviness of the tails increases from left to right in the plots.}}
\label{fig:pareto_a_infinite}
\end{figure}

\begin{figure}[!ht]
\begin{center}
\begin{tabular}{cc}
\includegraphics[width=84mm,clip=true]{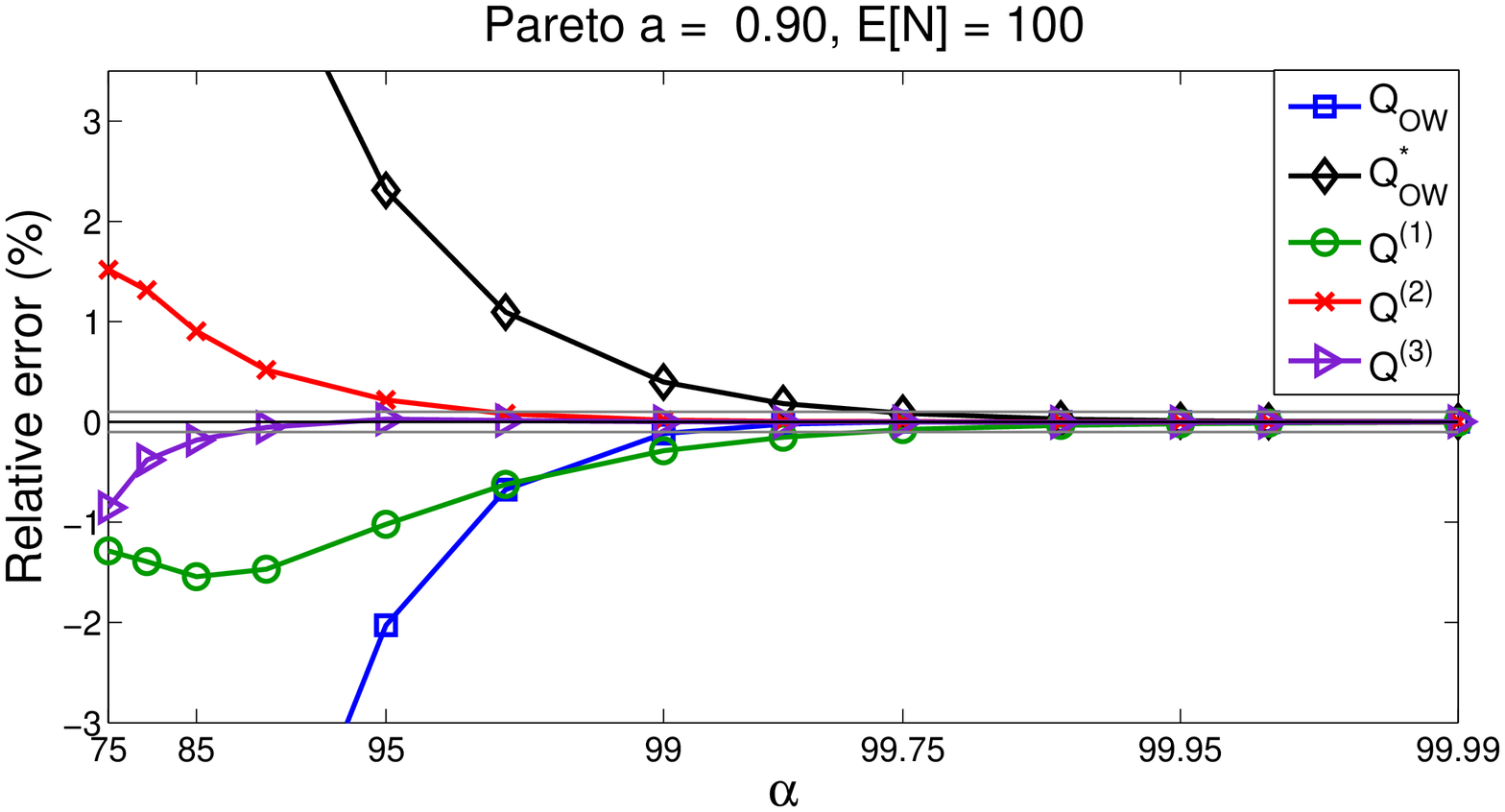} \\
\includegraphics[width=84mm,clip=true]{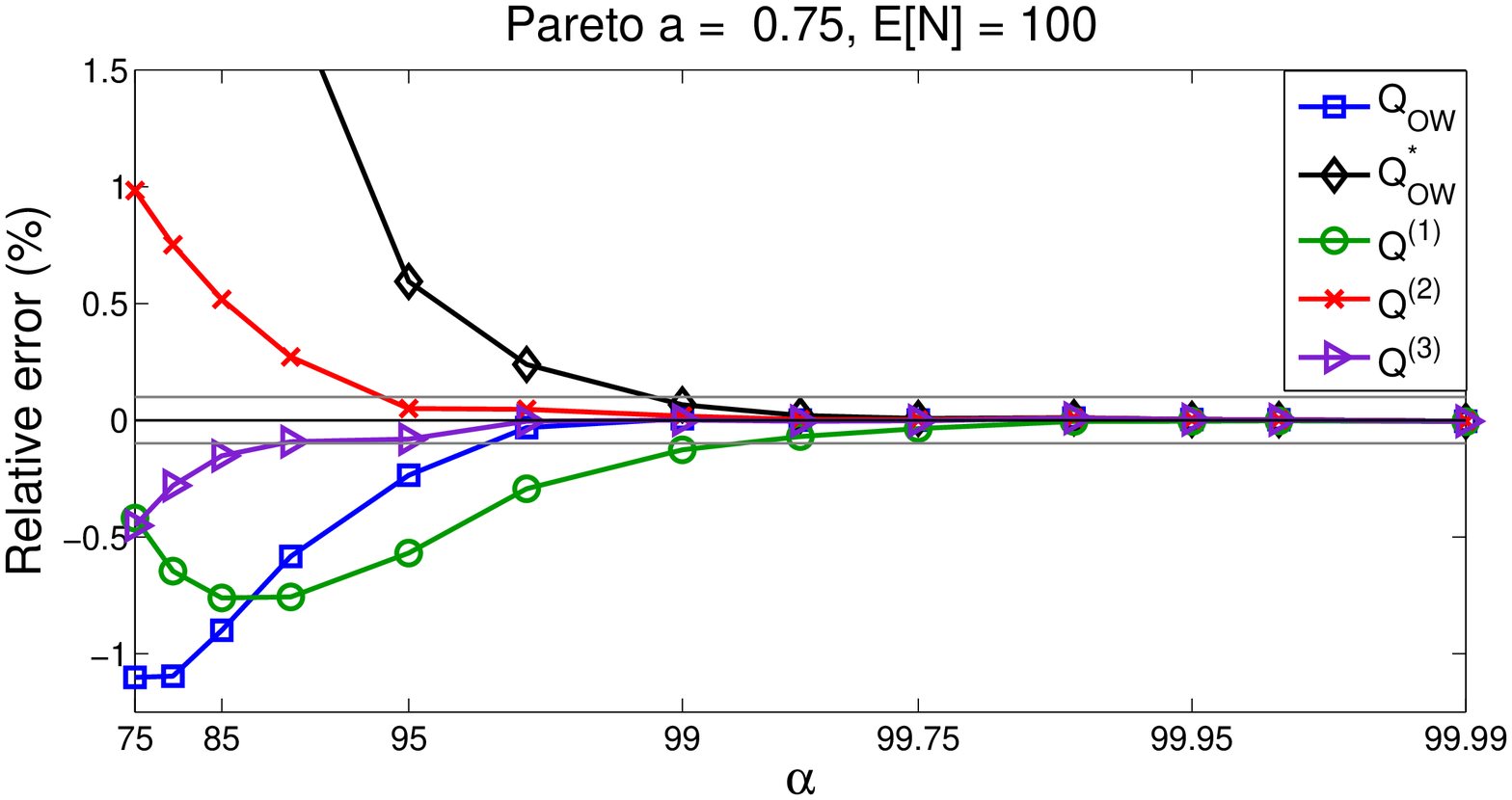} 
\end{tabular}
\end{center}
\caption{{\small Relative error for Poisson/Pareto   
as a function of $\alpha$ for $\lambda = 100$
and different values of $a$.}}
\label{fig:pareto_alpha_infinite}
\end{figure}

\begin{figure}[!ht]
\begin{center}
\begin{tabular}{cc}
\includegraphics[width=84mm,clip=true]{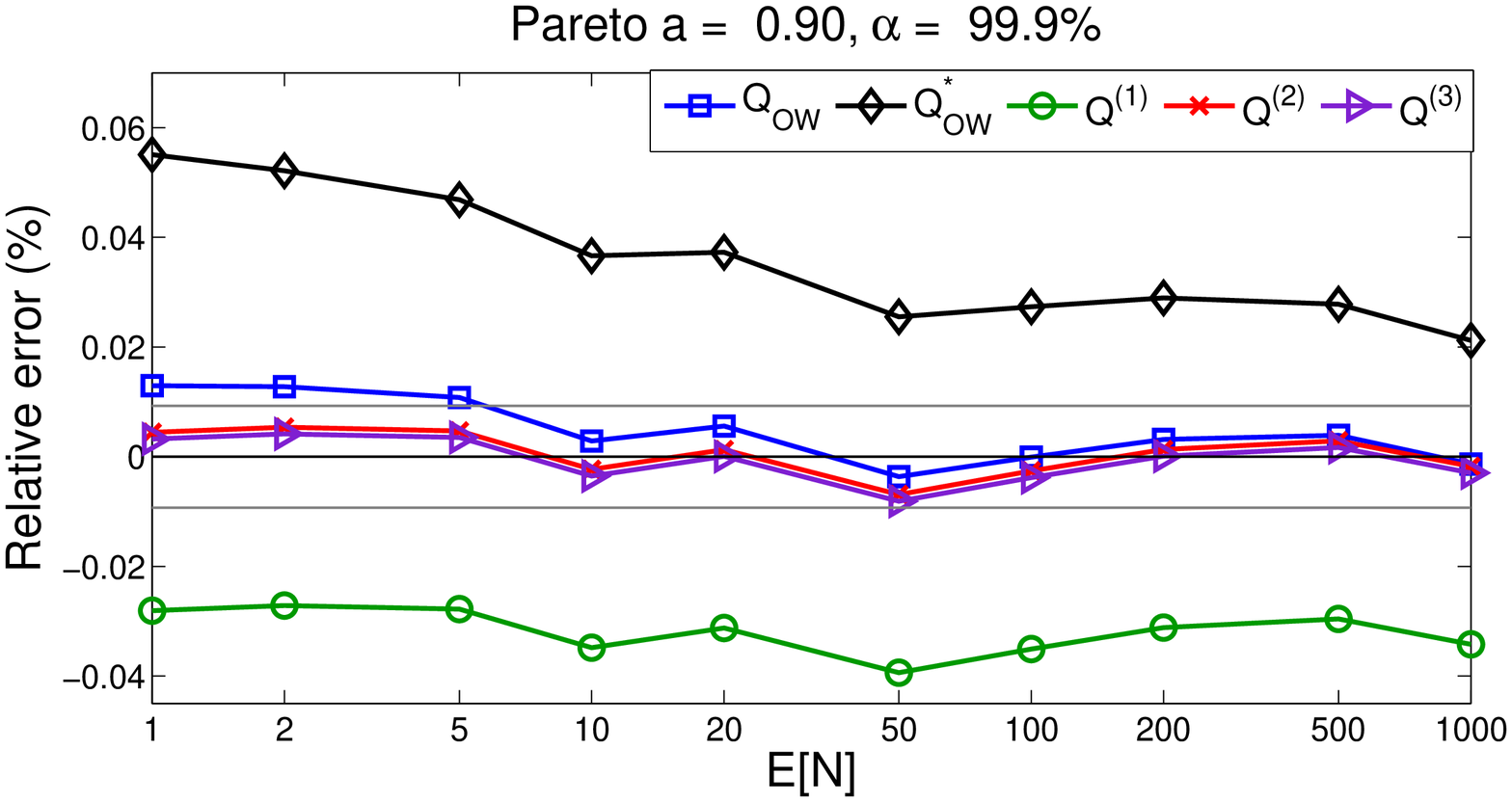} \\
\includegraphics[width=84mm,clip=true]{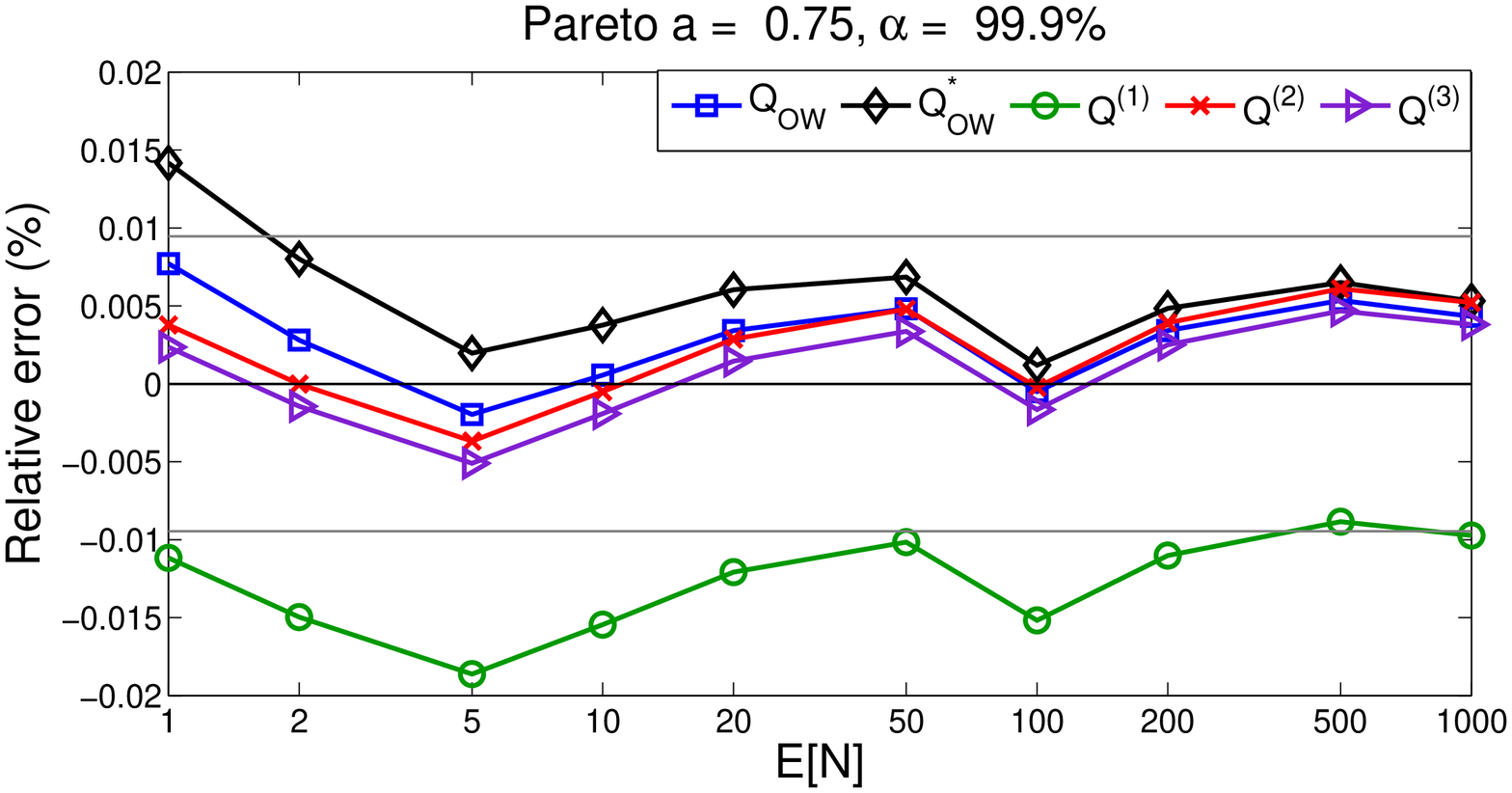} 
\end{tabular}
\end{center}
\caption{{\small Relative error for Poisson/Pareto  
as a function of $\EX[N]$ for $ \alpha = 99.9\% $ 
and different values of $a$.}}
\label{fig:pareto_lambda_infinite}
\end{figure}

We now evaluate the accuracy of the different approximations
for the percentiles of sums of random variables
that follow a Pareto distribution with infinite mean.
Figure \ref{fig:pareto_a_infinite} displays the 
relative error of the different approximations as
a function of $a$, the tail parameter of the Pareto distribution.
Figure \ref{fig:pareto_alpha_infinite} plots the relative
error as a function of $\alpha$ for two different values of $ a $.
Finally, the change in relative error as 
the average frequency $ \EX[N] $ varies is presented in
Figure \ref{fig:pareto_lambda_infinite}. 
In this regime all approximations are fairly accurate.
Between the second order asymptotic approximations,
$Q_{OW}$ is more accurate than $Q_{OW}^*$.

For high percentiles, the best results corresponds  to $Q^{(3)}$, the third order perturbative approximation. 
Beyond $\alpha = 0.90$ the errors of
this approximation are below the uncertainty of the 
Monte Carlo estimates.
The improvements with respect to the
standard approximations, $Q_{OW}$ or $Q_{OW}^*$,
are especially significant for values of $ a $ 
close to $1$.

\nopagebreak

\newpage

\subsubsection{Effective expansion parameter} \label{sssec:expansionParameter}

Equation  (\ref{EXPANSION}) has been derived using a purely formal 
expansion parameter $\epsilon$, which is eventually set to $1$. 
In this section we take advantage of the simple form of the Pareto 
distribution to identify the actual perturbative parameter of the expansion
for this type of random variables.
To this end, we analyze the leading contributions in the individual 
terms in the expansion for $ \alpha \rightarrow 1^{-}$. 
In terms of the parameter $\delta = (1-\alpha)$, the leading contributions for
$\delta \rightarrow 0^{+}$ and for all non-integer $a \neq \frac{1}{2}$ are
\begin{align}
\frac{Q_1}{N-1} \sim & -\frac{a}{a-1}  \left( \frac{\delta}{N}\right)^{1-1/a}
+ \ldots \nonumber\\ & +\frac{a}{a-1}  \left(\frac{\delta}{N}\right)^{0/a}  + \ldots \nonumber\\
\frac{Q_2}{N-1} \sim & -\frac{a(2a-1)}{a-2}  \left( \frac{\delta}{N}\right)^{1-1/a}  + \ldots \nonumber \\
& -\frac{a(a+1)}{(a-1)^2(a-2)} \left(\frac{\delta}{N}\right)^{1/a} + \ldots \nonumber\\
\frac{Q_3}{N-1} \sim &  -\frac{2a(a-1)(2a-1)}{a-3}  \left( \frac{\delta}{N}\right)^{1-1/a} + \ldots \nonumber\\ & +\frac{2a(a+1)^2(a+2)}{(a-1)^3(a-2)(a-3)}  \left(\frac{\delta}{N}\right)^{2/a}  + \ldots
\end{align}
The pattern that emerges is the following: up to order $Q_k$, with 
$k < a$, the terms  $ (\delta/N)^{(k-1)/a} $ dominate. 
Therefore, for $ k < a$,
$(\delta/N)^{1/a}$ can be interpreted as an expansion parameter.
For  $ k > a$ the terms  proportional to
$(\delta/N)^{1-1/a}$ dominate. Since these terms are independent of $k$, 
there is no longer a recognizable expansion parameter.
However, the prefactors, which depend on $a$,
become smaller as the order of the perturbative term increases. 
For $k=a$ both types of terms contribute. It is interesting 
to note that the dominance shifts precisely at
the order in which the moments cease to exist.

\section{Conclusions} \label{sec:conclusions}

Starting from a perturbative expansion for the percentile 
of a sum of two random variables we derive
a formal expansion for the percentile
of sum of $N$ independent random
variables. Assuming that, for sufficiently high percentiles,
the maximum dominates the sum, the expansion 
is carried around the percentile of the maximum in the sum.
This zeroth order term in the perturbative series
is similar to the single-loss approximation \cite{boecker+klueppeberg_2005_operational},
which can be derived from a first order asymptotic analysis
of the tails of sums of subexponential random variables
\cite{embrechts+veraverbeke_1982_estimates}.
The first order perturbative correction is similar to 
the mean-corrected single-loss formula for distributions
with finite mean \cite{boecker+sprittulla_2006_operational},
which can also be derived using higher order asymptotics. 
Higher order terms in the perturbative 
series are expressed in terms of right-truncated moments.
These censored moments are always finite, regardless of whether
the original uncensored distributions have finite or
divergent moments.
The perturbative series becomes more accurate for 
higher percentiles and heavier tails.  
From the empirical study carried out using
either exact results or Monte Carlo simulation, one
concludes that the perturbative approach 
is more accurate than 
previous approximate formulas proposed in the literature
\cite{sahay++_2007_operational,degen_2010_calculation,albrecher++_2010_higherOrder}.
Furthermore, the accuracy of the approximation can be improved by
including more terms in the perturbative series, up to a 
certain order. Beyond this order the approximation 
error generally increases. Another practical difficulty 
is the computational cost of the computations
of higher order terms.
Nonetheless, the third order approximation is
sufficiently accurate for the percentiles ($ 99-99.9 \%$), 
and the types of distributions that are used in practice in 
many fields of application, such as finance and insurance. 
As an extension of this research,
the perturbative analysis is being applied to sums
of random variables that are not identically distributed and may 
have dependencies. A more detailed analysis of the convergence of the 
perturbative series and the development of accurate approximations for lower
percentiles are also the subject of current investigation.

\section*{Acknowledgements}
A.S. acknowledges financial support from the Spanish 
\emph{Direcci\'on General de Investigaci\'on},
project TIN2010-21575-C02-02.

\newpage
\appendix
\section{Complete Bell polynomials} \label{app:bell}
The complete Bell polynomials (CBP) (named after Bell, \cite{BELL}) 
arise in many contexts, such as the n-times differentiation of a 
function (Fa\`a di Bruno formula) or to express the 
relationship between moments and cumulants in statistics.

Let $z(t)$ be an arbitrary function of $ t $
whose $k$-th derivative $z^{(k)}(t)=\frac{d^k}{dt^k}z(t)$ 
the complete Bell polynomial  of order $k$ is
\begin{equation} 
\label{A_BELL1}
B_k(z^{(1)}(t),\ldots,z^{(k)}(t)) = e^{-z(t)} \frac{d^k}{dt^k} e^{z(t)}
\end{equation} 
From this definition, the CBP can be shown to satisfy
\begin{equation}
\label{A_BELL2}
\exp{\left( \sum_{p=1}^{\infty} x_p \frac{t^p}{p!} \right)}  = 1 + \sum_{q=1}^{\infty} B_q \left( x_1 , \ldots , x_q \right)    \frac{t^q}{q!}.
\end{equation}
This expression provides a relationship between the power series expansion 
of the moment generating function and the cumulant generating function. 
In partucular
\begin{equation} 
\label{A_MOMENTSBELL}
\mu_q = B_q(\kappa_1,\ldots,\kappa_q),
\end{equation} 
where $\mu_q$ are the moments of a random variable and $\kappa_p$ its cumulants.

In this paper we use a centered version of the CBP, which is defined by 
$C_k(x_2,\ldots,x_k) \equiv B_k(0,x_2,\ldots,x_k)$.
In terms of $C_k(x_2,\ldots,x_k)$ the complete Bell polynomial of order $k$ is
\begin{equation} 
\label{A_CENTRALBELL}
B_k(x_1,\ldots,x_k) = \sum_{s=0}^k \binom{k}{s} x_1^s C_{k-s} (x_2,\ldots, x_{k-s})
\end{equation}

The CBP satisfy the following recursive formulae
\begin{equation} 
\begin{aligned}
k=0,     & \quad B_0 = 1 \quad , \quad C_0 = 1\\
k=1,     & \quad B_1(x_1) = x_1 \quad, \quad C_1 = 0\\
k \geq 2,& \quad B_k(x_1,\ldots,x_k) = \\
 & \quad x_k + \sum_{s=1}^{k-1} \binom{k - 1}{s - 1} x_s B_{k-s}(x_1,\ldots,x_{k-s}) \\
         & \quad C_k(x_2,\ldots,x_k) = \\
 & \quad x_k + \sum_{s=2}^{k-2} \binom{k - 1}{s - 1} x_s C_{k-s}(x_2,\ldots,x_{k-s}) \\
\end{aligned}
\end{equation}
There exists an alternative representation for the CBP, 
which is related to the structure of the partitions of a set of size $n$ 
\begin{equation}
\label{A_PARTITIONSBELL} 
B_k(x_1,\ldots,x_k) = \sum_{\Pt \in \PtsS(k)} \prod_{\elemPt \in \Pt}x_{|\elemPt|}
\end{equation}
where $\PtsS(k)$ is the set of all partitions of the set 
$\{1,\ldots,k\}$ (if $k=0$, $\PtsS(k)$ contains one empty set) 
and $|\mathtt{b}|$ denotes the number of elements in set $\mathtt{b}$.

\section{Explicit formulas for particular frequency distributions} \label{app:frequency}
In this section we provide explicit formulas for the 
first terms in the perturbative series when the
number of terms in the sum is distributed as  
a Poisson or as a negative binomial.
\subsubsection{Poisson distribution}
We consider the particular case where $N$, the number
of terms in the sum (\ref{eq:sum_randomFrequency}) 
follows a Poisson distribution with parameter $\lambda = \EX[N]$
\begin{equation}
p_n =  \frac{1}{n!} \lambda^n e^{-\lambda}.
\end{equation}
The moment generating function is
\begin{equation} 
\begin{aligned}
\mathcal{M}_N(s) &= \exp{\left( \lambda(e^s-1) \right)}.
\end{aligned}
\end{equation}
From this we derive
\begin{equation} 
\begin{aligned}
\lambda_0(x)        &= \exp{\left( \lambda (F(x)-1)  \right)} \lambda f(x)\\
\lambda_1(x)        &= \lambda F(x) \lambda_0(x) \\
\lambda_2(x)        &= (1+\lambda F(x)) \lambda_1(x).
\end{aligned}
\end{equation}
The first three terms of the perturbative expansion are
\begin{equation}
\begin{aligned}
Q_0 &= F^{-1} \left( \frac{\log{\alpha}}{\lambda} + 1 \right) \\
Q_1 &= \left( \lambda + \log{\alpha}\right) \EX \left[ L\vert L<Q_0\right] \\
Q_2 &= -\frac{1}{\lambda_0(Q_0)}\partial_x \left(\lambda_1(x) (\kappa_2(x)+\kappa_1(x)^2) \right)_{x=Q_0} \nonumber \\
    &= -\left(\lambda f(Q_0)+\frac{f'(Q_0)}{f(Q_0)}\right) (\log{\alpha}+\lambda) \EX [ L^2 | L<Q_0 ] \\
    & \hspace*{0.35cm}-\lambda f(Q_0) Q_0^2,
\end{aligned}
\end{equation}
where, in the last step, we have used the identity
\begin{equation}
\partial_x \mu_p(x) = \frac{f(x)}{F(x)}\left( x^p - \mu_p(x)\right).
\end{equation}
for the censored moments $\mu_p(x) \equiv \EX [L^p | L<x ]$.
 
\subsubsection{The negative binomial distribution}
The probability mass function
of the negative binomial distribution with parameters $(p, r)$ is 
\begin{equation}
	p_n =  \binom{n+r-1}{n} p^r(1-p)^n.
\end{equation}
Setting $q \equiv 1-p$ The moment generating function is
\begin{equation} 
	\mathcal{M}_N(s) = p^r\left[ 1-q e^s\right]^{-r}.
\end{equation}
In terms of $\xi(x) = 1-qF(x)$ we have 
\begin{equation} 
\begin{aligned}
\lambda_0(x)        &= p^r {\xi(x)}^{-r-1} q r f(x)\\
\lambda_1(x)        &= {\xi(x)}^{-1}q(1+r)F(x) \lambda_0(x)\\
\lambda_2(x)        &= {\xi(x)}^{-1}\left[1+q(1+r)F(x)\right]\lambda_1(x)
\end{aligned}
\end{equation}
The first three terms in the perturbative expansion are
\begin{equation} 
\begin{aligned}
Q_0 &= F^{-1} \left( \frac{1-p{\alpha}^{-1/r}}{q} \right) \\
Q_1 &= (1+r)\left( \frac{\alpha^{1/r}}{p} - 1 \right) \EX \left[ L\vert L<Q_0\right] \\
Q_2 &=-\frac{r+1}{h^3} \Bigg[ \\
    &  \EX \left[ L^2\vert L<Q_0\right] h(1-h)\left( q(r+2)f(Q_0) + h\frac{f'(Q_0)}		
       {f(Q_0)} \right) \\
 	&  + \EX \left[ L\vert L<Q_0\right]^2 (1-h)^2\left( q(r+3)f(Q_0) + h\frac{f'(Q_0)}    
 	   {f(Q_0)} \right) \\ 
 	&  + \EX \left[ L\vert L<Q_0\right] 2 Q_0 q h(1-h) f(Q_0) +  {Q_0}^2 q h^2 f(Q_0) \Bigg]
\end{aligned}
\end{equation}

where $q\equiv1-p$ and $h \equiv p\alpha^{-1/r} $.

\section{Recursive formulas for the perturbative series}
\label{RECURSIVE_FORMULAE}
The objective of this section is to derive recursive 
expressions for the terms in the perturbative expansion 
of high quantiles of $Z=X+\epsilon Y$. 
These expressions are better suited for the numerical 
computation of the series than the  
expressions derived in section \ref{sec:perturbativeExpansion}. 
\\
The starting point is (\ref{BASIC_EQUATION}).
By defining the function 
\begin{equation}
	\phi(s,x) \equiv  f_X(x) \mathcal{M}_{Y|X}(s\vert x),
\end{equation}
where  
\begin{equation}
\mathcal{M}_{Y|X}(s\vert x) = \int_{-\infty}^{\infty} \, dy \, e^{sy} f_{Y|X}(y\vert x)
\end{equation}
is the moment
generation function of $Y$ conditional on $X$,
and the operator
\begin{equation}
\Omega_{\epsilon} \equiv 
\left( e^{\left(\delta Q  - \epsilon\partial_s \right)\partial_x} - 1 \right) \partial_x^{-1},
\end{equation}
(\ref{BASIC_EQUATION}) can be written as 
\begin{equation}
	\left. \Omega_{\epsilon} \phi(s,x) \right|_{s = 0, x = Q_0} = 0.
\end{equation}
By defining the operators $ \left\{ \Omega^{(n)} \equiv \frac{\partial^n \Omega_{\epsilon}}{\partial \epsilon^n}\vert_{\epsilon=0}, \ n \ge 0 \right\}$, 
the terms of the perturbative expansion can be obtained by solving the equations 
\begin{equation}
\label{MASTER}
\left.\Omega^{(n)} \phi(s,x)\right|_{s = 0, x = Q_0} = 0\, , \, \, \, \text{for all }\, \, \, n \geq 0. 
\end{equation}
The sequence of operators $\Omega^{(n)}$ has the recurrence relation
\begin{eqnarray}
\label{MASTER_RECURSIVE}
\Omega^{(0)} &=& 0  \nonumber\\
\Omega^{(1)} &=& \widetilde{\partial}_s \nonumber\\
\Omega^{(k)} & =& Q_{k} + \Omega^{(k-1)}\widetilde{\partial}_s \partial_x + \displaystyle{\sum_{i=1}^{k-2}} \binom{k-1}{i} Q_{k-i}\Omega^{(i)}\partial_x  
\end{eqnarray}
for $k \geq 2$ and with $\widetilde{\partial}_s \equiv Q_1 - \partial_s$. 
Expressing each operator $\Omega^{(n)}$ in the form
\begin{equation}
\Omega^{(k)} = \sum_{i=0}^n \sum_{j=0}^n \omega_{i,j}^{(k)} \widetilde{\partial}_s^i\partial_x^j
\end{equation}
(\ref{MASTER_RECURSIVE}) can be expressed as
recursion relations for the coefficients
\begin{align}
\omega^{(k)}_{0,0} &= Q_k & \nonumber\\
\omega^{(k)}_{i,0} &= 0     \nonumber\\
\omega^{(k)}_{0,j} &= \displaystyle{\sum_{l=\max(1,j-1)}^{k-2}}\binom{k-1}{l}Q_{k-l} \omega_{0,j-1}^{(l)} \nonumber\\
\omega^{(k)}_{i,j} &= \displaystyle{\sum_{l=\max(1,i,j-1)}^{k-2}}\binom{k-1}{l}Q_{k-l} \omega_{i,j-1}^{(l)} + \omega_{i-1,j-1}^{(k-1)},
\end{align} 
for $i,j \geq 1$. Finally, the terms in the perturbative series 
can be derived from (\ref{MASTER}) as
\begin{eqnarray}
\label{Q_EQUATIONS}
Q_n = -\frac{1}{\phi^{(0,0)}} \sum_{i=0}^n \sum_{j=1}^n \omega_{i,j}^{(n)} \phi^{(i,j)}
\end{eqnarray}
where
\begin{equation}
\phi^{(i,j)} \equiv  
\left. \widetilde{\partial}_s^i\partial_x^j\phi(s,x)\right|_{s = 0, x = Q_0}.
\end{equation}
The remainder of this appendix is devoted to the derivation of 
explicit recursive formulas for the quantities $\phi^{(i,j)}$ 
of the perturbative expansion for sums of $N$ 
independent random variables, $Z_N = \sum_{n=1}^N L_n$. These independent
rv's are identically distributed according to $F(l)$ (density $f(l)$).

\subsection{Deterministic $N$}
Consider the case of sums of $N$ iidrv's, with $N$ fixed. 
In this case, the expansion around the maximum of the terms in 
the sum is characterized by
\begin{align}
	f_X(x) &= NF(x)^{N-1}f(x) \nonumber \\ 
	\mathcal{M}_{Y|X}(s\vert x) &= \mathcal{M}_L(s\vert x)^{N-1}.
\end{align}
Therefore
\begin{equation}
\phi(s,x) = Nf(x)F(x)^{N-1} \mathcal{M}_L(s\vert x)^{N-1}.
\end{equation}
To make the notation more compact, the following definitions are 
used in the derivation  
\begin{align}
\mathcal{M}_{Y|X}^{(i,j)} &\equiv  \left. \partial_s^i \partial_x^j \mathcal{M}_L^{N-1}(s\vert x)\right|_{s = 0, x = Q_0}\nonumber\\
k_i^{(j)} & \equiv  \left. \partial_x^j \kappa_i(x)\right|_{x = Q_0} = 
\left. \partial_s^i \partial_x^j \log \mathcal{M}_L(s\vert x)\right|_{s = 0, x = Q_0}
\nonumber\\
{m}^{(j)}_i & \equiv   \left. \partial_x^j \mu_i(x)\right|_{x = Q_0} =
\left. \partial_s^i \partial_x^j  
\mathcal{M}_L(s\vert x)\right|_{s = 0, x = Q_0}\nonumber\\ 
\tilde{F}^{(j)} & \equiv \left. \partial_x^j  \log F(x)\right|_{x = Q_0} \nonumber\\
\tilde{f}^{(j)} & \equiv \left. \partial_x^j  \log f(x)\right|_{x = Q_0}\, , 
\end{align}
where 
\begin{equation}
\mathcal{M}_L(s\vert x) = \int_0^x dl \, e^{sl}  \, \frac{f(l)}{F(x)} 
\end{equation}
is the generating function of the censored moments 
of the individual terms in the sum with censoring
threshold $x$.

Using these expressions and definitions, the 
coefficients in (\ref{Q_EQUATIONS}) are
\begin{equation}
\phi^{(i,j)} = \sum_{l=0}^i\sum_{k=0}^j (-1)^{i-l} \binom{j}{k} \binom{i}{l} 
Q_1^l  \mathcal{M}_{Y|X}^{(i-l,j-k)} \left. \partial_x^k f_X(x) \right|_{x=Q_0}.	
\end{equation}
The derivatives of the conditional moment generating function
$\mathcal{M}_{Y|X}(s\vert x)$ evaluated at $s=0$ and $x = Q_0$
can be computed using the recursion 
\begin{align}
\mathcal{M}_{Y|X}^{(0,j)} &= \delta_{0,j} \nonumber\\
\mathcal{M}_{Y|X}^{(i,j)} &=(N-1) \displaystyle{\sum_{l=0}^{i-1}\sum_{k=0}^{j}} \binom{i-1}{l}\binom{j}{k} \mathcal{M}_{Y|X}^{(l,k)} k_{i-l}^{(j-k)} 
\end{align}
for $j \geq 0$, $i \geq 1$ and with $\delta_{i,j}$ the Kronecker delta. 
To evaluate this expression one 
needs the derivatives of the censored cumulants evaluated at
$Q_0$.  These can be computed using the recursion 
\begin{align}
k_0^{(j)} &=  0 \nonumber \\
k_i^{(j)} &= m_i^{(j)}- \displaystyle{\sum_{l=1}^{i-1} \sum_{k=0}^{j}} \binom{i-1}{l} \binom{j}{k} m_l^{(k)} k_{i-l}^{(j-k)}  
\end{align} 
for $j \geq 0$, $i \geq 1$. 
Finally, the derivatives of the censored moments evaluated at $Q_0$
are given by the recursion
\begin{equation}
m_i^{(j)} = \displaystyle{\sum_{k=0}^{j-1}\binom{j-1}{k}}  
\tilde{F} ^{(1+k)} \left( \left. \partial_x^{j-1-k} x^i \right|_{x = Q_0} - m_i^{(j-1-k)} \right) , 
\end{equation}
for $j \geq 1$. 
Besides $m_i^{(0)}$, the censored moments with threshold $x=Q_0$, 
the remaining terms in the calculation, namely, the derivatives of 
logarithm of the severity $\tilde{F}^{(j)}$ and of the density of 
the maximum $ \partial_x^k f_X(x) =  \partial_x^k \left( N f(x) F(x)^{N-1}\right) $ 
can be readily computed from the derivatives of the severity CDF, 
also via recursion. For instance
\begin{eqnarray}
\tilde{F}^{(j)} = \frac{ \left[ \partial_x^j F(x) - \displaystyle{\sum_{k=1}^{j-1} \binom{j-1}{k}\partial_x^k F(x)\tilde{F}^{(j-k)}} \right]_{x = Q_0} }{F(Q_0)}.
\end{eqnarray}

\subsection{Random $N$}

In this case 
\begin{align}
	\mathcal{M}_{Y|X}(s|x) &= \mathbb{E}\left[ e^{SY} \big\vert X=x \right] \nonumber\\
	&= \sum_{n=0}^{\infty} \mathcal{M}_{Y|X,N}(s|x,n) \frac{ f_{X_n}(x) p_n}{f_X(x)} \nonumber\\
	&= \sum_{n=0}^{\infty} \mathcal{M}_L(s|x)^{n-1} \frac{ f_{X_n}(x) p_n}{f_X(x)} 
\end{align}
with $p_n\equiv\mathbb{P}[N=n]$.
In terms of these quantities
\begin{equation}
	\phi(s,x) = \mathbb{E} \left[ f_{X_N}(x)\mathcal{M}_L(s|x)^{N-1} \right]. 
\end{equation}
The coefficients in (\ref{Q_EQUATIONS}) are then given by
\begin{align}
\phi^{(i,j)} &= \sum_{l=0}^i \binom{i}{l} Q_1^l(-1)^{i-l} \partial_s^{i-l} \partial_x^j \mathbb{E}\Big[ f_{X_N}(x) \mathcal{M}_L^{N-1}(s,x) \Big]_{s=0}^{x=Q_0} \nonumber \\ 
& =  \sum_{l=0}^i \binom{i}{l} Q_1^l(-1)^{i-l} \xi^{(i-l,j)}_0,	
\end{align}
where 
\begin{equation}
\xi^{(i,j)}_a = \partial_s^i \partial_x^j \mathbb{E}\Big[ (N-1)^a f_{X_N}(x) \mathcal{M}_L^{N-1}(s,x) \Big]_{s=0}^{x=Q_0}. 
\end{equation}
These quantities have the recursion
\begin{align}
\xi^{(0,j)}_a &=  \partial_x^j \mathbb{E}\Big[ (N-1)^a f_{X_N}(x) \Big]_{x=Q_0} = \left. \partial_x^j \lambda_a(x) \right|_{x = Q_0} \nonumber \\
\xi^{(i,j)}_a &= \displaystyle{\sum_{l=0}^{i-1}\sum_{k=0}^{j}} \binom{i-1}{l}\binom{j}{k} \xi^{(l,k)}_{a+1} k_{i-l}^{(j-k)}  i \geq 1,\, j \geq 0
\end{align}
where the coefficients $\lambda_a(x)$ have been defined in (\ref{LAMBDA_DEFINITION}). Their values at $x=Q_0$ can be computed using equation (\ref{LAMBDA_FROM_MGF}) in terms of the derivatives of the moment generating function. To obtain the derivatives $\lambda_a^{(k)} \equiv 
\left. \partial_x^k \lambda_a(x) \right|_{x = Q_0} $ in the previous equation, the following recursion can be used
\begin{equation}
	\lambda_a^{(k)} = \sum_{l=0}^{k-1} \binom{k-1}{l} \Big(  \tilde{f}^{(l+1)} \lambda_a^{(k-l-1)}  + \tilde{F}^{(l+1)}\lambda_{a+1}^{(k-l-1)} \Big).
\end{equation}
The remaining elements in the calculation 
(moments, censored cumulants and their derivatives etc.) 
are computed as in the case with deterministic $N$.

\section{Derivation of higher order asymptotic approximations} \label{app:relatedWork}
In this section we present the derivations of the single-loss approximation
and higher order corrections that have been given in the literature. 
\cite{omey+willekens_1986_second,omey+willekens_1987_second,sahay++_2007_operational,degen_2010_calculation,barbe+mccormick_2005_asymptotic,barbe++_2007_asymptotic,barbe+mccormick_2009_asymptotic,albrecher++_2010_higherOrder} 
\subsection{Second order approximation by Omey and Willekens
\cite{omey+willekens_1986_second,omey+willekens_1987_second}} 
It is possible to derive corrections to the single-loss approximation by using 
the second order behavior of the tail probability of subordinate distributions  \cite{omey+willekens_1986_second,omey+willekens_1987_second}. 
These references are also the basis for the analysis presented in \cite{sahay++_2007_operational,degen_2010_calculation}.

For the case in which the mean is finite, the second order approximation 
for the tail distribution of the sum is 
\cite{omey+willekens_1987_second}
\begin{equation}
1- G(x) \sim \mathtt{E}\left[N \right] \left(1-F(x) \right) 
+ \mathtt{E}\left[N(N-1)\right]  \mu_L f(x) \quad \quad x \rightarrow \infty.
\end{equation}
From this, it is possible to derive a nonlinear equation
for a second order approximation of
$Q \equiv G^{-1}(\alpha) $, the percentile of the sum at the probability level $\alpha$
\begin{equation} \label{eq:A_singleLossMeanCorrected}
Q \approx 
F^{-1}\left[1 - \frac{1-\alpha}{\mathtt{E}\left[N\right]} + 
\left(\frac{\mathtt{E}\left[N^2 \right]}{\mathtt{E}\left[N \right]} -1 \right)  
\mu_L f \left(Q \right) \right]. 
\end{equation} 
This nonlinear equation can be solved numerically.

A closed-form expression that is similar to the correction by the mean 
proposed in \cite{boecker+sprittulla_2006_operational} is obtained using
an approximate  solution of
\begin{equation} \label{eq:A_secondOrderPerturbative}
1- \alpha \sim \mathtt{E}\left[N \right] \left(1-F(Q) \right) 
+ \epsilon \mathtt{E}\left[N(N-1)\right]  \mu_L f(Q) 
\end{equation}
where the parameter $\epsilon = 1$ has been introduced
to order the terms in a perturbative expansion of the solution
\begin{equation}
Q = {Q'}_0  + \epsilon {Q'}_1 + \ldots
\end{equation}
Expanding  (\ref{eq:A_secondOrderPerturbative}) up to first order in 
$\epsilon$ we obtain
\begin{eqnarray*}
1- \alpha  & = &  
\mathtt{E}\left[N \right] \left(1-F({Q'}_0  + \epsilon {Q'}_1 + \ldots) \right) \nonumber \\ 
& & + \epsilon \mathtt{E}\left[N(N-1)\right]  \mu_L f({Q'}_0  + \epsilon {Q'}_1 + \ldots) \nonumber \\
1- \alpha  & = &  \mathtt{E}\left[N \right] \left(1-F({Q'}_0) \right) \nonumber \\ 
& & - \epsilon 
 \Big(\mathtt{E}\left[N \right] f({Q'}_0) {Q'}_1 -
 \mathtt{E}\left[N(N-1)\right]  \mu_L f({Q'}_0) \Big) \nonumber\\
 & & + {\cal O}(\epsilon^2) \nonumber
\end{eqnarray*}
Identifying terms of the same order,
\begin{align}
1- \alpha  & =   \mathtt{E}\left[N \right] \left(1-F({Q'}_0) \right) \quad \Longrightarrow \nonumber \\
& \quad \quad  {Q'}_0  = F^{-1}\left(1- \frac{1-\alpha}{\mathtt{E}\left[N \right]} \right),  \\
0 & = \left(\mathtt{E}\left[N \right] {Q'}_1 -
 \mathtt{E}\left[N(N-1)\right]  \mu_L \right) f({Q'}_0) 
\quad \Longrightarrow \nonumber\\ 
& \quad \quad  {Q'}_1  =  \left( \frac{\mathtt{E}\left[N^2\right]}{\mathtt{E}\left[N \right]}  -1 \right) \mu_L,  
\end{align}
which provides a good approximation to the solution provided that
$f({Q'}_0) >0$ and ${Q'}_1 \ll {Q'}_0$.
Therefore, the approximate solution of (\ref{eq:A_secondOrderPerturbative}) with
$\epsilon = 1$ is 
\begin{equation} \label{eq:A_meanCorrection}
Q \approx F^{-1}\left(1 - \frac{1-\alpha}{\mathtt{E}\left[N\right]} \right) + \left(\mathtt{E}\left[N \right] + (D-1) \right) \mu_L,
\end{equation}
where $ D = \mathtt{Var}\left[N \right]/\mathtt{E}\left[N \right]$ is the index of dispersion
($D = 1$ for the Poisson distribution and $ D > 1$ for the negative binomial distribution).
The first term in (\ref{eq:A_meanCorrection}) is the single-loss formula. 
The second term is a correction that involves the mean.

Similar approximate formulas can be given for  
the case of distributions $F(L)$ with infinite mean
and whose corresponding density is regularly varying $ f(L) \in RV_{-(1+a)}$
using the results of \cite{omey+willekens_1986_second} 
\begin{align}
1- G(x) & \sim \mathtt{E}\left[N \right] \left(1-F(x) \right) + c_a  
\mathtt{E}\left[N (N-1) \right] \mu_{F}(x) f(x) \nonumber \\ 
& \text{for} \quad x \rightarrow \infty,
\end{align}
where
\begin{equation}
\mu_{F}(x) \equiv \int_0^x ds (1-F(s)) = 
(1-F(x)) x + F(x) \mathtt{E}\left[L | \le x \right],
\end{equation}
and 
\begin{equation}
c_a = \left\{ 
\begin{array}{ll}
 1 & a = 1 \\
(1- 1/a) {\frac{\left[\Gamma(1-a) \right]^2}{2 \Gamma(1-2a)}} 
& a < 1
\end{array}
\right. .
\end{equation}

In this case a second order approximation of
$Q \equiv G^{-1}(\alpha) $ can be obtained from 
\begin{equation}
Q  \approx  F^{-1}\left(1- \frac{1- \alpha}{\mathtt{E}\left[N \right]} + c_a  
\left( \frac{\mathtt{E}\left[N^2\right]}{\mathtt{E}\left[N \right]}  -1 \right) \mu_F(Q) f(Q) \right). 
\end{equation}
Again, this nonlinear equation can be solved numerically
using, for example, an iterative scheme. Alternatively, 
an approximate closed-form expression can be obtained 
by means of a perturbative scheme analogous to the finite mean case
\begin{eqnarray} \label{eq:A_infiniteMeanCorrection}
Q  & \approx & {Q'}_0 
+ c_{a} \left(\mathtt{E}\left[N \right] + (D-1) \right) \mu_F({Q'}_0), \\
{Q'}_0 & = & F^{-1}\left(1 - \frac{1-\alpha}{\mathtt{E}\left[N\right]} \right) \\
\mu_F({Q'}_0)  & = & \int_0^{{Q'}_0} \left(1-F(s) \right) ds \nonumber\\
 &=& \frac{1-\alpha}{\mathtt{E}\left[N\right]} {Q'}_0  + 
\left(1 - \frac{1-\alpha}{\mathtt{E}\left[N\right]} \right) \mathtt{E}\left[L | L \le {Q'}_0 \right],\nonumber\\
\end{eqnarray}

\subsection{Asymptotic expansion by Barbe and McCormick 
\cite{barbe+mccormick_2005_asymptotic,barbe++_2007_asymptotic,barbe+mccormick_2009_asymptotic}}
This section uses the approximations for the distribution of  
sums of independent random variables with heavy tails derived in
\cite{barbe+mccormick_2005_asymptotic,barbe++_2007_asymptotic,barbe+mccormick_2009_asymptotic}. 
For simplicity, we assume that  the
number of terms in the sum are sampled from a Poisson distribution.
Assuming that the first $m$ moments of the 
variables in the sum are finite
\begin{align}
1 - G(x) & = \lambda \exp\left\{ \lambda 
\sum_{i=1}^m \frac{(-1)^i}{i!} \mu_L^{[i]} 
\partial_x^i \right\}
\left[1-F(x) \right] \nonumber\\
 & + \mathcal{O} \left(h^m(x) \left[1-F(x) \right] \right), 
\end{align}
where $ h(x) = f(x)/(1-F(x)) $ and 
$ \mu_L^{[i]} \equiv E[L^i] = \int_0^{\infty} dx \, f(x) x^i $
is the $i$th moment of $L$.
For $m=0$, the single-loss approximation is recovered. 
The first order approximation ($m=1$) is 
\begin{equation} \label{eq:babe+mccormick}
1 - G(x) \approx \lambda e^{-\lambda \mu_L \partial_x}
\left[1-F(x) \right].  
\end{equation}
In \cite{barbe++_2007_asymptotic,barbe+mccormick_2009_asymptotic} the authors
proceed by preforming a Taylor expansion of the right-hand side of (\ref{eq:babe+mccormick}). 
Here, we derive an exact formula by realizing
that the Taylor expansion can be resummed. This resummation
results in a translation of the argument of $F$
\begin{equation} 
1 - G(x) \approx \lambda  \left[1-F(x - \lambda \mu_L) \right].  
\end{equation}
Therefore, the first order approximation  to the 
$\alpha$ percentile of $G$  yields the correction by the mean 
\begin{equation}
Q \approx F^{-1}\left(1 - \frac{1-\alpha}{\lambda} \right) + \lambda \mu_L,
\end{equation}
also in this derivation.
The second order approximation for $G$ can also be expressed in terms 
of an integral over a diffusion kernel
\begin{align}
1 - G(x) & \approx \lambda \Bigg[1 - \int_{0}^{\infty} \, dz \, 
\frac{1}{\sqrt{2 \pi \lambda \mu_L^{[2]}}} \times \nonumber\\
& \quad \quad \quad  \exp\left\{- \frac{\left(z- x +\lambda \mu_L \right)^2}{2\lambda \mu_L^{[2]}} \right\} 
F(z) \Bigg].  
\end{align}
The corresponding second order approximation ($m=2$)
for $Q$, the $\alpha$ percentile
of $G$ is the solution of the nonlinear equation
\begin{align}
1 - \alpha & = \lambda \Bigg[1 - \int_{0}^{\infty} \, dz \, 
\frac{1}{\sqrt{2 \pi \lambda \mu_L^{[2]}}}\nonumber\\
& \quad \quad \quad \exp\left\{- \frac{\left(z- Q +\lambda \mu_L \right)^2}{2\lambda \mu_L^{[2]}} \right\} 
F(z) \Bigg].  
\end{align}

\subsection{Asymptotics with a shifted argument}
The results of this section are based on the expansion 
for $G$ derived  in \cite{albrecher++_2010_higherOrder} using only evaluations
of $F$ at different arguments
\begin{equation}
1-G(x) \approx \mathtt{E}\left[N \right] \left(\xi_1  F(x - k_1) + \ldots + \xi_m  F(x - k_m) \right)
\end{equation}
for some constants $\xi_1,\ldots,\xi_m$, $k_1,\ldots,k_m$.
Assuming that the  first $m$ moments of $F$ are finite, these
constants are the solution of the system of
equations 
\begin{eqnarray}
\sum_{j=1}^m \xi_j & = & 1 \nonumber \\
\sum_{j=1}^m \left(c_i - \mathtt{E}\left[ N \right] k_j^i\right)\xi_j & = & 1; 
i = 1,\ldots,m,
\end{eqnarray}
where $c_i = \mathtt{E}\left[N \left(X_1 + \ldots + X_N \right)^i \right]; \ \text{for} \ \ge 0$.  There is some freedom in the choice $k_1,\ldots,k_m$. 
In \cite{albrecher++_2010_higherOrder} the authors 
propose to determine the values of these parameters by enforcing
the constraints
\begin{equation}
\sum_{j=1}^m \left(c_{m+i} - \mathtt{E}\left[ N \right] k_j^{m+i}\right)\xi_j  =  0, \quad \text{for} \quad 
i = 1,\ldots,m-1.
\end{equation}
Therefore, the approximation of order $m$ is obtained by solving the set 
of nonlinear equations
\begin{equation}
\sum_{j=1}^m  \xi_j k_j^{i} = \tilde{c}_{i}, \quad \text{for} \quad 
i = 0,\ldots,2m-1.
\end{equation}
where $\tilde{c}_{i} = c_{i} / \mathtt{E}\left[N \right] \ \text{for} \ i \ge 0$.

For $m=1$
\begin{eqnarray}
\xi_1  =  1, \quad \quad
k_1    =  \tilde{c}_1 = \frac{c_1}{\mathtt{E}\left[N\right]} = 
\frac{\mathtt{E}\left[N(N-1) \right]}{\mathtt{E}\left[N\right]} \mu_L, 
\end{eqnarray}
which yields the first order approximation
\begin{equation}
1-G(x) \approx \mathtt{E}\left[N \right] F(x - \frac{\mathtt{E}\left[N(N-1) \right]}{\mathtt{E}\left[N\right]} \mu_L). 
\end{equation}
The $\alpha$ percentile of $G$ in this approximation is 
the single-loss formula corrected by the mean (\ref{eq:A_meanCorrection}).

\section{Approximations to high percentiles of sums of L\'evy 
iidrv's} \label{app:Levy}
The exact $\alpha$ quantile for the sum of $N$ Levy iidrv's 
with parameters $(\mu=0,c)$ is
\begin{equation}
Q = \frac{c}{2} N^2 \left[\hbox{erf}^{-1}(\delta)\right]^{-2},
\end{equation}
where $ \delta = 1-\alpha$.
High percentiles can be approximated as
\begin{equation}
Q \approx \frac{2c}{\pi} N^2 \Bigg[ \frac{1}{\delta^2} -\frac{\pi}{6} -\frac{\pi^2}{120}\delta^2 + \mathcal{O}(\delta^4) \Bigg], \quad \delta \rightarrow 0^+.
\end{equation}
For the L\'evy distribution the approximation to the quantiles (\ref{eq:OW_infiniteMean}) is
\begin{equation}
Q_{OW} = \frac{c}{2} \left[\hbox{erf}^{-1}\left( \frac{\delta}{N} \right) \right]^{-2}. 
\end{equation}
For high percentiles, this approximation is of the form
\begin{equation}
Q_{OW} \approx \frac{2c}{\pi} \Bigg[ \frac{N^2}{\delta^2} -\frac{\pi}{6} + \mathcal{O}\left(\left(\frac{\delta}{N}\right)^2\right) \Bigg], \quad \delta \rightarrow 0^+.
\end{equation}
Similarly, it is possible derive the high-percentile 
approximations of the perturbative expansion coefficients  
\begin{equation}
\begin{aligned}
Q_0 \approx &\frac{2 N^2 c}{\pi} \Bigg[ \frac{1}{\delta^2} & -\frac{N-1}{N}\frac{1}{\delta} &+ \frac{(N-1)(N-5)-2 \pi}{12 N^2} 
\hspace*{-0.35cm}&+ \mathcal{O}(\delta) \Bigg]\\
Q_1 \approx &\frac{2 N^2 c}{\pi} \Bigg[ & \frac{N-1}{N}\frac{1}{\delta} 
& - \frac{(N - 1)(N + \pi - 3)}{2 N^2} 
\hspace*{-0.35cm}&+ \mathcal{O}(\delta) \Bigg]\\
Q_2 \approx &\frac{2 N^2 c}{\pi} \Bigg[ &  & - \frac{(N - 1)(N + 1)}{6 N^2} 
&\hspace*{-0.35cm}+ \mathcal{O}(\delta) \Bigg]\\
Q_3 \approx &\frac{2 N^2 c}{\pi} \Bigg[ &  & - \frac{(N-1)(N-2)}{5N^2} 
&\hspace*{-0.35cm}+ \mathcal{O}(\delta) \Bigg].
\end{aligned}
\end{equation}

\bibliographystyle{unsrt}     
\bibliography{beyondSingleLoss}

\begin{thebibliography}{10}

\bibitem{nadarajah_2008_review}
Saralees Nadarajah.
\newblock A review of results on sums of random variables.
\newblock {\em Acta Applicandae Mathematicae}, 103:131--140, 2008.

\bibitem{cohen_1972_tail}
J.W. Cohen.
\newblock On the tail of the stationary waiting-time distribution and limit
  theorem for {M/G/1} queue.
\newblock {\em Annales de l'Institut Henri Poincar{\'e}}, 8:255--263, 1972.

\bibitem{fay++_2006_modeling}
Gilles {Fa{\"y}}, B{\'a}rbara Gonz{\'a}lez-Ar{\'e}valo, Thomas Mikosch, and
  Gennady Samorodnitsky.
\newblock Modeling teletraffic arrivals by a poisson cluster process.
\newblock {\em Queueing Systems}, 54:121--140, 2006.

\bibitem{embrechts++_1997_modelling}
P.~Embrechts, C.~Kl{\"u}ppelberg, and T.~Mikosch.
\newblock {\em Modelling extremal events for insurance and finance}.
\newblock Applications of mathematics. Springer, 1997.

\bibitem{mcneil2005}
A.J. McNeil, R.~Frey, and P.~Embrechts.
\newblock {\em Quantitative risk management: concepts, techniques and tools}.
\newblock Princeton series in finance. Princeton University Press, 2005.

\bibitem{frachot++_2001_loss}
Antoine Frachot, Pierre Georges, and Thierry Roncalli.
\newblock Loss distribution approach for operational risk.
\newblock {\em Working paper, Groupe de Recherche Op\'erationnelle, Cr\'edit
  Lyonnais, France (2001)}, 2001.

\bibitem{embrechts++_2003_quantifying}
Paul Embrechts, Hansj{\"o}rg Furrer, and Roger Kaufmann.
\newblock Quantifying regulatory capital for operational risk.
\newblock {\em Derivatives Use, Trading {\&} Regulation}, 9(3):217--233, 2003.

\bibitem{panjer_2006_operational}
H.H. Panjer.
\newblock {\em Operational Risk: Modeling Analytics}.
\newblock Wiley, New York, 2006.

\bibitem{carrilloMenendez+suarez_2012_robust}
Santiago Carrillo-Men{\'e}ndez and Alberto Su{\'a}rez.
\newblock Robust quantification of the exposure to operational risk: Bringing
  economic sense to economic capital.
\newblock {\em Computers {\&} OR}, pages 792--804, 2012.

\bibitem{klugman++_2004_loss}
S.~A. Klugman, H.~H. Panjer, and G.~E. Willmot.
\newblock {\em Loss Models: from Data to Decisions, 2nd edn.}
\newblock Wiley, 2004.

\bibitem{dupire1998monte}
B.~Dupire.
\newblock {\em Monte Carlo: methodologies and applications for pricing and risk
  management}.
\newblock Risk Books, 1998.

\bibitem{asmussen++_2000_rare}
{S\o ren} Asmussen, Klemens Bingswanger, and Bjarne {H\o jgaard}.
\newblock Rare events simulation for heavy-tailed distributions.
\newblock {\em Bernoulli}, 6:303--322, 2000.

\bibitem{asmussen+kroese_2006_improved}
{S\o ren} Asmussen and Dirk~P. Kroese.
\newblock Improved algorithms for rare event simulation with heavy tails.
\newblock {\em Advances in Applied Probability}, 38(2):545--558, 2006.

\bibitem{boecker+klueppeberg_2005_operational}
K.~B{\"o}cker and C.~Kl{\"u}ppelberg.
\newblock Operational var: a closed-form approximation.
\newblock {\em Risk}, December:90--93, 2005.

\bibitem{goldie+klueppelberg_1998_subexponential}
Charles~M. Goldie and Claudia Kl{\"u}ppelberg.
\newblock Subexponential distributions.
\newblock In Robert~J. Adler, R.~Feldman, and M.~S. Taqqu, editors, {\em A
  Practical Guide to Heavy Tails: Statistical Techniques and Applications},
  pages 435--459. Birkh{\"a}user, Boston, MA, 1998.

\bibitem{foss+etal_subexponential}
S.~Foss, D.~Korshunov, and S.~Zachary.
\newblock {\em An Introduction to Heavy-Tailed and Subexponential
  Distributions}.
\newblock Springer Series in Operations Research and Financial Engineering.
  Springer, 2011.

\bibitem{reiss2007statistical}
R.D. Reiss and M.~Thomas.
\newblock {\em Statistical analysis of extreme values: with applications to
  insurance, finance, hydrology and other fields}.
\newblock Birkh{\"a}user, 2007.

\bibitem{asmussen2003applied}
S.~Asmussen.
\newblock {\em Applied probability and queues}.
\newblock Applications of mathematics. Springer, 2003.

\bibitem{tsourti+panaretos_2004_extreme_telefraffic}
Zoi Tsourti and John Panaretos.
\newblock Extreme value analysis of teletraffic data.
\newblock {\em Computational Statistics {\&} Data Analysis}, 45:85--103, 2004.

\bibitem{crovella+taqqu+bestavros_1998_heavyTailed}
Mark~E. Crovella, Murad~S. Taqqu, and Azer Bestavros.
\newblock {\em Heavy-tailed probability distributions in the World Wide Web},
  pages 3--25.
\newblock Birkhauser Boston Inc., Cambridge, MA, USA, 1998.

\bibitem{resnick_2007_heavy}
S.I. Resnick.
\newblock {\em Heavy-tail phenomena: probabilistic and statistical modeling}.
\newblock Number v. 10 in Springer series in operations research. Springer,
  2007.

\bibitem{omey+willekens_1986_second}
E.~Omey and E.~Willekens.
\newblock Second-order behaviour of the tail of a subordinated probability
  distribution.
\newblock {\em Stochastic Processes and their Applications}, 21:339 -- 353,
  1986.

\bibitem{omey+willekens_1987_second}
E.~Omey and E.~Willekens.
\newblock Second-order behaviour of distributions subordinate to a distribution
  with finite mean.
\newblock {\em Communications in Statistics. Stochastic Models}, 3(3):311 --
  342, 1987.

\bibitem{grubel_1987_subordinated}
Rudolf Grubel.
\newblock On subordinated distributions and generalized renewal measures.
\newblock {\em Annals of Probability}, 15(1):394--415, 1987.

\bibitem{sahay++_2007_operational}
Anupam Sahay, Zailong Wan, and Brian Keller.
\newblock Operational risk capital: asymptotics in the case of heavy-tailed
  severity.
\newblock {\em The journal of operational risk}, 2(2):61--72, 2007.

\bibitem{degen_2010_calculation}
M.~Degen.
\newblock The calculation of minimum regulatory capital using single-loss
  approximations.
\newblock {\em Journal of Operational Risk}, 5(4):3--17, 2010.

\bibitem{barbe+mccormick_2005_asymptotic}
Philippe Barbe and William~P. McCormick.
\newblock Asymptotic expansions of convolutions of regularly varying
  distributions.
\newblock {\em Journal of the Australian Mathematical Society}, 78(3):339--371,
  2005.

\bibitem{barbe++_2007_asymptotic}
Ph. Barbe, W.~P. McCormick, and C.~Zhang.
\newblock Asymptotic expansions for distributions of compound sums of random
  variables with rapidly varying subexponential distribution.
\newblock {\em Journal of Applied Probability}, 44(3):670--684, 2007.

\bibitem{barbe+mccormick_2009_asymptotic}
P.~Barbe and W.P. McCormick.
\newblock {\em Asymptotic expansions for infinite weighted convolutions of
  heavy tail distributions and applications}.
\newblock Memoirs of the American Mathematical Society. American Mathematical
  Society, 2009.

\bibitem{albrecher++_2010_higherOrder}
Hansj{\"o}rg Albrecher, Christian Hipp, and Dominik Kortschak.
\newblock Higher-order expansions for compound distributions and ruin
  probabilities with subexponential claims.
\newblock {\em Scandinavian Actuarial Journal}, 2010(2):105--135, 2010.

\bibitem{boecker+sprittulla_2006_operational}
K.~B{\"o}cker and J.~Sprittulla.
\newblock Operational var: meaningful means.
\newblock {\em Risk}, December:96--98, 2006.

\bibitem{nolan_2012_stable}
J.~P. Nolan.
\newblock {\em Stable Distributions - Models for Heavy Tailed Data}.
\newblock Birkhauser, Boston, 2012.
\newblock In progress, Chapter 1 online at
  academic2.american.edu/$\sim$jpnolan.

\bibitem{nadarajah_2006_distribution}
Saralees Nadarajah and M.~Masoom Ali.
\newblock The distribution of sums, products and ratios for lawrance and
  lewis's bivariate exponential random variables.
\newblock {\em Computational Statistics {\&} Data Analysis}, 50(12):3449 --
  3463, 2006.

\bibitem{GOURIEROUX}
C.~Gourieroux, J.P. Laurent, and O.~Scaillet.
\newblock Sensitivity analysis of values at risk.
\newblock {\em Journal of Empirical Finance}, 7:225--245, 2000.

\bibitem{WILDE}
R.~Martin and T.~Wilde.
\newblock Unsystematic credit risk.
\newblock {\em Risk}, 15(11):123--128, 2002.

\bibitem{Blum_1970}
M.~Blum.
\newblock On the sums of independently distributed {Pareto} variates.
\newblock {\em SIAM Journal on Applied Mathematics}, 19(1):191--198, July 1970.

\bibitem{chistyakov_1964_theorem}
V.~P. Chistyakov.
\newblock A theorem on sums of independent positive random variables and its
  applications to branching random processes.
\newblock {\em Theory of Probability and its Applications}, 9(4):640--648,
  1964.

\bibitem{embrechts+veraverbeke_1982_estimates}
P.~Embrechts and N.~Veraverbeke.
\newblock Estimates for the probability of ruin with special emphasis on the
  possibility of large claims.
\newblock {\em Insurance: Mathematics and Economics}, 1(1):55 -- 72, 1982.

\bibitem{Bingham1987}
N.~H. Bingham, C.~M. Goldie, and J.~L. Teugels.
\newblock {\em Regular Variation}.
\newblock Cambridge University Press, 1987.

\bibitem{basel2_2006_international}
{Basel Committee on Banking Supervision}.
\newblock {\em International Convergence of Capital Measurement and Capital
  Standards. A Revised Framework}.
\newblock June 2006.

\bibitem{BELL}
E.T. Bell.
\newblock Exponential polynomials.
\newblock {\em Annals of Mathematics}, 35:258--277, 1934.

\end{thebibliography}

\end{document}